\documentclass[%
superscriptaddress,
amsmath,amssymb,
aps,
prb,
]{revtex4-2}

\usepackage{graphicx}% Include figure files
\usepackage{dcolumn}% Align table columns on decimal point
\usepackage{bm}% bold math
\usepackage{hyperref}% add hypertext capabilities
\usepackage{booktabs}% nice tables
\usepackage{multirow}
\usepackage{xcolor} % for showing changes to reviewers

\begin{document}

	\title{Dome-like pressure-temperature phase diagram of the cooperative Jahn--Teller distortion in NaNiO$_2$}

	\author{Liam A. V. Nagle-Cocco}
	\altaffiliation{Present address: Stanford Synchrotron Radiation Lightsource, SLAC National Accelerator Laboratory, Menlo Park, California 94025, United States.}
	\email{Email: lnc@stanford.edu. Corresponding author.}
	\affiliation{Cavendish Laboratory, University of Cambridge, JJ Thomson Avenue, Cambridge, CB3 0HE, United Kingdom.}
	
	\author{James M. A. Steele}
	\affiliation{Cavendish Laboratory, University of Cambridge, JJ Thomson Avenue, Cambridge, CB3 0HE, United Kingdom.}
	\affiliation{Yusuf Hamied Department of Chemistry, University of Cambridge, Cambridge, CB2 1EW, United Kingdom.}
	
	\author{Shiyu Deng}
	\altaffiliation{Present address: Institute Laue Langevin, BP 156, 38042, Grenoble, France.}
	\affiliation{Cavendish Laboratory, University of Cambridge, JJ Thomson Avenue, Cambridge, CB3 0HE, United Kingdom.}
	
	\author{Xiaotian Zhang}
	\affiliation{Cavendish Laboratory, University of Cambridge, JJ Thomson Avenue, Cambridge, CB3 0HE, United Kingdom.}
	
	\author{Dominik Daisenberger}
	\affiliation{Diamond Light Source, Diamond House, Harwell Science and Innovation Campus, Didcot, OX11 ODE, Oxfordshire, United Kingdom.}
	
	\author{Annalena R. Genreith-Schriever}
	\affiliation{Yusuf Hamied Department of Chemistry, University of Cambridge, Cambridge, CB2 1EW, United Kingdom.}
	
	\author{Siddharth S. Saxena}
	\affiliation{Cavendish Laboratory, University of Cambridge, JJ Thomson Avenue, Cambridge, CB3 0HE, United Kingdom.}
	\affiliation{British Management University Tashkent, 35 Bobur Mirza Street, Tashkent, Uzbekistan.}
	\affiliation{Kazakh-British Technical University, 59 Tole Bi Str., 050000, Almaty, Kazakhstan.}
	
	\author{Clare P. Grey}
	\affiliation{Yusuf Hamied Department of Chemistry, University of Cambridge, Cambridge, CB2 1EW, United Kingdom.}
	
	\author{Si\^an E. Dutton}
	\email{Email: sed33@cam.ac.uk. Corresponding author.}
	\affiliation{Cavendish Laboratory, University of Cambridge, JJ Thomson Avenue, Cambridge, CB3 0HE, United Kingdom.}

	\date{\today}% It is always \today, today,
	%  but any date may be explicitly specified

	\begin{abstract}
		NaNiO$_2$ is a Ni$^{3+}$-containing layered material consisting of alternating triangular networks of Ni and Na cations, separated by octahedrally-coordinated O anions. At ambient pressure, it features a collinear Jahn--Teller distortion below $T^\mathrm{JT}_\mathrm{onset}\approx480$\,K, which disappears in a first-order transition on heating to $T^\mathrm{JT}_\mathrm{end}\approx500$\,K, corresponding to the increase in symmetry from monoclinic to rhombohedral. 
		It was previously studied by variable-pressure neutron diffraction [ACS Inorganic Chemistry 61.10 (2022): 4312-4321] and found to exhibit an increasing $T^\mathrm{JT}_\mathrm{onset}$ with pressure up to $\sim$5\,GPa. 
		In this work, powdered NaNiO$_2$ was studied \textit{via} variable-pressure synchrotron x-ray diffraction up to pressures of $\sim$67\,GPa at 294\,K and 403\,K. 
		Suppression of the collinear Jahn--Teller ordering is observed \textit{via} the emergence of a high-symmetry rhombohedral phase, with the onset pressure occurring at $\sim$18\,GPa at both studied temperatures. 
		Further, a discontinuous decrease in unit cell volume is observed on transitioning from the monoclinic to the rhombohedral phase. 
		These results taken together suggest that in the vicinity of the transition, application of pressure causes the Jahn--Teller transition temperature, $T^\mathrm{JT}_\mathrm{onset}$, to decrease rapidly. 
		We conclude that the pressure-temperature phase diagram of the cooperative Jahn--Teller distortion in NaNiO$_2$ is dome-like. 
	\end{abstract}

	\maketitle

	\section{Introduction}
	
	In octahedrally-coordinated transition metal ions, the Jahn--Teller (JT) effect arises when there is a degenerate occupancy of $d$ orbitals, and often results in an octahedral elongation~\cite{jahn1937stability,van1939jahn,nagle2024van}. 
	Within crystal structures, such distortions may couple to one another and to macroscopic strain, resulting in correlated axes of elongation; this is termed a cooperative JT distortion. 
	JT distortions are important in battery materials, as they impact ionic mobility and can lead to structural transitions on cycling if the JT-active ion undergoes a change in oxidation state to a non-JT-active state~\cite{li2016jahn}. 
	The JT effect is also important in high-temperature superconductivity~\cite{fil1992lattice,keller2008jahn} and colossal magnetoresistance~\cite{merten2019magnetic}. 
	
	Several previous works have studied the effect of pressure on the JT distortion using variable-pressure diffraction~\cite{aasbrink1999high,loa2001pressure,waskowska2001cumn2o4,zhou2008breakdown,zhou2011jahn,aguado2012pressure,mota2014dynamic,caslin2016competing,zhao2016pressure,collings2018disorder,bostrom2019high,li2020size,bhadram2021pressure,lawler2021decoupling,scatena2021pressure,ovsyannikov2021structural,nagle2022pressure,bostrom2024pressure}. 
	Most of these have shown that pressure reduces the magnitude of the JT distortion as a consequence of the elongated bond being more compressible than the shorter bonds. 
	Some have shown that application of pressure results in a structural phase transition related to the JT distortion. 
	For instance, LaMnO$_3$ loses its cooperative JT distortion at $\sim$11\,GPa~\cite{zhou2008breakdown} before proceeding to metallise at $\sim$32\,GPa~\cite{loa2001pressure}, and ZnMn$_2$O$_4$ has a structural phase transition at $\sim$23\,GPa which is described as either an elongation$\rightarrow$compression transition~\cite{aasbrink1999high} or a spin--crossover-induced JT suppression~\cite{choi2006electronic}. 
	
	\begin{figure*}[!t]
		\includegraphics[scale=1]{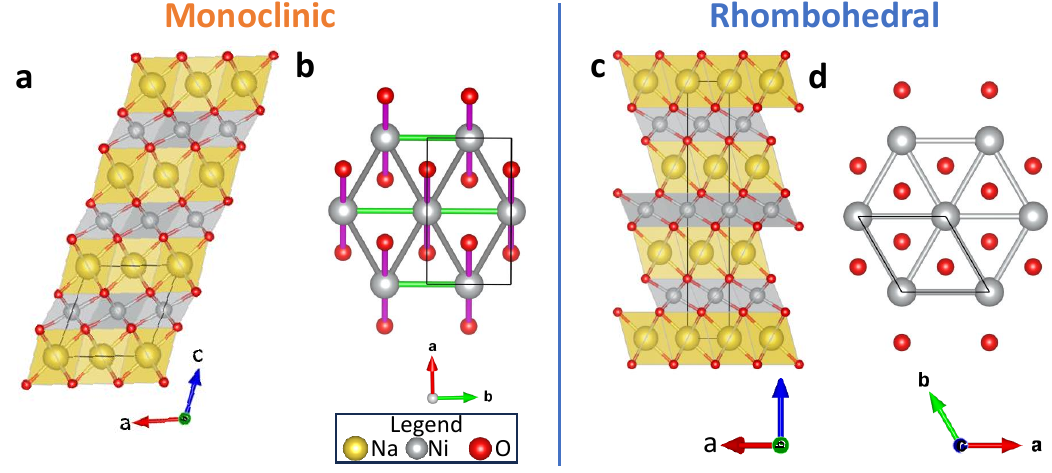}
		\caption{
			Ambient-pressure crystal structures of (a,b) $C2/m$ monoclinic and (c,d) $R\bar{3}m$ rhombohedral NaNiO$_2$. 
			(a) and (c) show the unit cells. 
			(b) and (d) show the Ni-Ni distances (grey for non-elongated; green for elongated) and JT-elongated Ni-O distances (purple). 
			In the rhombohedral structure there are no elongated Ni-Ni or Ni-O bonds. 
		}
		\label{NaNiO2-structures.pdf}
	\end{figure*}
	
	NaNiO$_2$ is a material which is of interest for studying collinear cooperative JT distortions on a layered triangular framework. 
	It consists of alternating layers of NiO$_6$ octahedra and Na$^{+}$ cations [Figure~\ref{NaNiO2-structures.pdf}], with triangular metal networks within each layer~\cite{dyer1954alkali}. 
	At ambient pressure, it has a rhombohedral ($R\bar{3}m$) structure at high-temperature, but at room-temperature up to $\sim$480\,K it is monoclinic ($C2/m$) with a collinear cooperative JT distortion, a result of the low-spin $d^7$ Ni$^{3+}$ cations. 
	The transition from monoclinic$\rightarrow$rhombohedral is a first-order transition~\cite{dick1997structure,chappel2000study,sofin2005new,nagle2024displacive}, which locally corresponds to a displacive transition~\cite{nagle2024displacive}, meaning that the loss of cooperative JT distortion also corresponds to a disappearance of local JT distortion~\cite{radin2020order,genreith2024jahn}. 
	NaNiO$_2$ has been studied as a battery cathode material~\cite{vassilaras2012electrochemical,han2014structural,wang2017unravelling,sada2024mitigating,steele2025structural}, and the broader class of layered nickelates are critically-important materials for rechargeable Li- and Na-ion batteries~\cite{booth2021perspectives}.
	
	We recently performed a variable-pressure neutron diffraction study on the layered nickelate NaNiO$_2$,  which measured isothermal variable-pressure diffraction at 290\,K, 460\,K, and 490\,K up to $\sim$6\,GPa~\cite{nagle2022pressure}. 
	%To our knowledge, this was the first attempt to study the JT distortion using both pressure \textit{and} temperature, with prior studies only studying as a function of one of these two parameters. 
	The primary finding of that study was that application of pressure increases the JT transition temperature within the studied temperature and pressure ranges. 
	We used this finding to propose a tentative pressure-temperature phase diagram~\cite{nagle2022pressure}. 
	
	A limitation of neutron diffraction is that large sample volumes ($\sim$100s of mg) are required, which limits the accessible pressure range. 
	In this study, we report results from a new diffraction experiment we have performed. In this experiment, we used a diamond anvil cell (DAC) with a far higher accessible pressure range, and investigated the structural changes in NaNiO$_2$ using x-ray radiation from a synchrotron source. 
	We substantially extend the pressure-temperature phase diagram proposed in our previous work and show that the region of the phase diagram in which there is a cooperative JT distortion is dome-like in shape. 
	This dome-like phase diagram is analogous to that obtained for superconductivity under pressure in some materials~\cite{knebel2006coexistence,torikachvili2008pressure,pan2015pressure,wang2023two}. 
	
	\section{Methods}
	
	%\subsection{Sample synthesis.} 
	
	\textit{Sample synthesis.} 
	Samples were prepared by solid state synthesis. 
	Na$_2$O$_2$ (Alfa Aesar; 95\%) and NiO (Alfa Aesar; 99.995\%) were mixed and pelletised in a 1.05:1 molar ratio of Na:Ni, with excess Na to account for Na-loss during heating. 
	Sample was heated to 973\,K for 70 hrs in a tube furnace under constant flow of O$_2$. 
	To prevent reaction with moisture, the sample was stored and handled in an inert Ar-atmosphere. 
	The same sample was previously studied in Ref.~\cite{nagle2022pressure}, where laboratory x-ray diffraction and electron microscopy were also presented, and Ref.~\cite{nagle2024displacive}.
	
	%\subsection{Synchrotron x-ray diffraction.}
	
	\textit{X-ray diffraction.} 
	Variable-pressure x-ray diffraction was performed using the I15 instrument~\cite{anzellini2018laser} at Diamond Light Source ($\lambda= 0.35421$\,\AA{}), using an aliquot of NaNiO$_2$ in a LeToullec-style 4-pin membrane Diamond Anvil Cell with a rhenium gasket. 
	NaNiO$_2$ was placed in the sample chamber with sample dimensions of approximately $50\times50\times20$\,$\mu$m. 
	The sample chamber also contained a micron-sized gold particle and a micron-sized ruby to act as pressure gauges. 
	A Ne pressure medium (grade 5; purity $>=99.999$\%; Air Products and Chemicals Inc) was used. 
	Temperature control was achieved using BETSA K-ring resistive heating, with temperature measured using K-type thermocouples.
	
	\textit{Diffraction data analysis.} 
	The integration of 2D area diffraction patterns was performed using Dioptas~\cite{prescher2015dioptas}. 
	1D integrated diffraction data were then analysed using the software package \textsc{topas 7}~\cite{coelho2018topas} using sequential Pawley refinement~\cite{pawley1981unit}. 
	An order-12 Chebyshev polynomial was used to fit the background profile, and a pseudo-Voigt function~\cite{thompson1987rietveld} was used to model Bragg peaks. 
	Axial divergence~\cite{cheary1998axial} is refined for the phases except for Au. 
	
	\textit{Pressure calibration.} 
	In this manuscript, claimed pressures were calibrated based on the Au unit cell volume. 
	These pressures are consistent with those obtained from the ruby. 
	For a full description of the pressure calibration, see Appendix 1. 

	\section{Results}
	
	\begin{figure*}%[t]
		\includegraphics[scale=1.0]{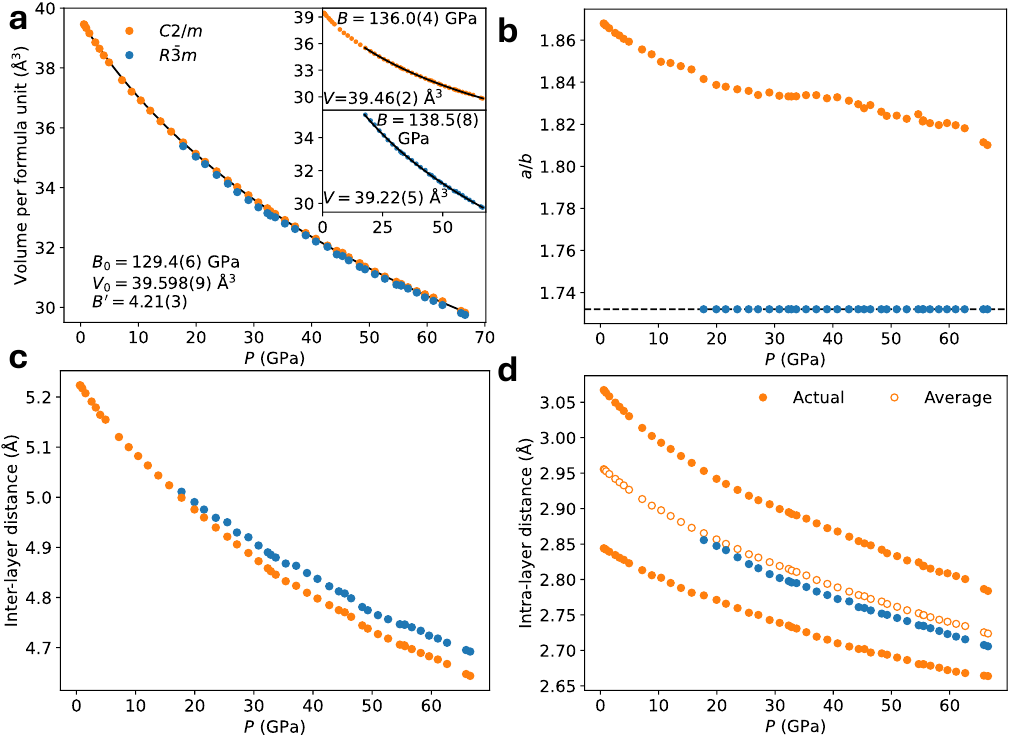}
		\caption{
			The pressure-dependence of the lattice parameters for the 294\,K isotherm. 
			(a) Unit cell volume per formula unit for the rhombohedral (blue) and monoclinic (orange) phases. Black line is a 3rd-order Birch-Murnaghan fit to the unit cell volume data for the monoclinic phase throughout the measured pressure range, along with associated parameters. 
			Insets show the monoclinic (upper) and rhombohedral (lower) unit cell volumes fit with a 2nd-order Birch-Murnaghan phase within the two-phase region only; note that in this range $B'\approx4$ so it is fixed to 4 for the fitting. 
			(b)	Ratio of the monoclinic $a$ and $b$ distances for the two phases of NaNiO$_2$, showing the reduction of magnitude of monoclinicity of NaNiO$_2$. 
			The dashed black horizontal line occurs at $a/b=\sqrt{3}$ for the complete loss of monoclinic distortion. 
			(c) inter-layer distances, $c/3$ for the rhombohedral phase and $c\cdot\sin{(\beta)}$ for the monoclinic phase, and (d) intra-layer metal-metal distances, $a$ for the rhombohedral phase, and $a/\sqrt{3}$ and $b$ for the monoclinic phase. Open circles indicate the average of the two distances. 
		}
		\label{LPs_with_pressure}
	\end{figure*}

	We studied an aliquot of powdered NaNiO$_2$ in two DACs as a function of pressure at 294\,K up to 66.58(4)\,GPa and at 403\,K up to 21.556(13)\,GPa. At the lowest measured pressures, it exists solely in the cooperatively JT-distorted monoclinic ($C2/m$) structure. 
	On compression to pressures beyond that previously studied in Ref.~\cite{nagle2022pressure}, there emerge a new set of peaks indexable to the same rhombohedral ($R\bar{3}m$) space group which exists at high-temperatures and ambient pressure in NaNiO$_2$, i.e. a phase that does not have a JT distortion in its average structure. 
	The rhombohedral phase emerges at a pressure between $15.68(4)<P^\mathrm{JT}_\mathrm{onset}$(GPa)$<17.77(3)$ at 294\,K, and between $14.175(9)<P^\mathrm{JT}_\mathrm{onset}$(GPa)$<17.992(11)$ at 403\,K. 
	Both the monoclinic and rhombohedral phases then co-exist throughout the measured pressure range all the way to 66.58(4)\,GPa at 294\,K, indicating that this is a first-order transition analogous to the ambient-pressure, variable-temperature transition~\cite{sofin2005new,nagle2024displacive}, although we do not observe the completion of the transition and can assume that $P^\mathrm{JT}_\mathrm{end}>66.58(4)$\,GPa at 294\,K and $P^\mathrm{JT}_\mathrm{end}>21.556(13)$\,GPa at 403\,K.

\begin{figure*}[t]
		\includegraphics[scale=1.0]{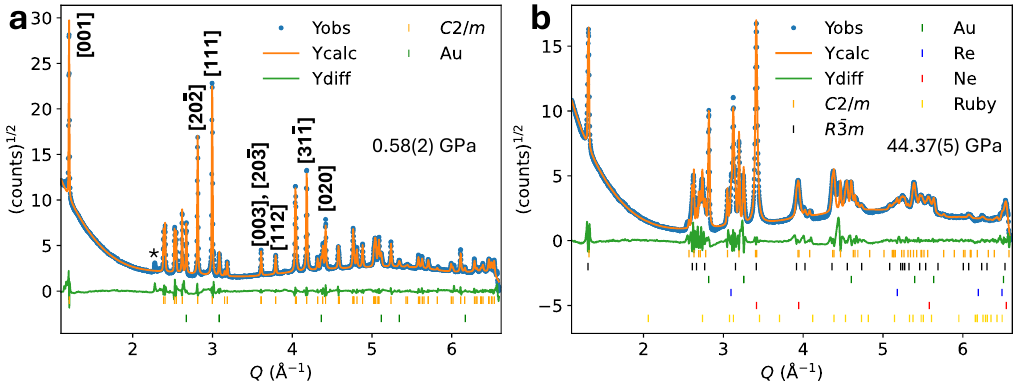}
		\caption{
			Representative Pawley refinements of NaNiO$_2$ at 294\,K, at (a) 0.58(2)\,GPa in the monoclinic regime and (b) 44.37(5)\,GPa in the biphasic regime. 
			The y-axis is the square root of the intensity, and the difference pattern is $y_\mathrm{diff}=\sqrt{y_\mathrm{obs}} - \sqrt{y_\mathrm{calc}}$. 
			The peak marked with an asterisk in (a) is unassigned, but it is clear from the distribution of intensity in the area detector pattern that it is not due to NaNiO$_2$.
		}
		\label{PawleyRefinementsFig}
	\end{figure*}

	At the lowest pressures, the only phase present in the diffraction pattern besides monoclinic NaNiO$_2$ was cubic ($Fm\bar{3}m$) gold, due to the pressure marker. 
	In addition to the onset of the pressure-induced monoclinic$\rightarrow$rhombohedral phase transition, on application of pressure more phases gradually became present due to the constricting of the sample chamber relative to the diameter of the beam. 
	At the highest pressure, it was necessary to include rhenium ($P63/mmc$) due to the gasket, ruby ($R\bar{3}c$) due to the pressure marker, and crystalline Ne ($Fm\bar{3}m$) due to crystallisation of the pressure medium; see Figure~\ref{PawleyRefinementsFig} for example diffraction patterns. 
	
	In the 294\,K variable-pressure isotherm, there is a small peak on the low-$2\theta$ (high-$d$) side of the monoclinic [001] peak. This has been attributed to a small amount of moisture exposure in NaNiO$_2$~\cite{vassilaras2012electrochemical,han2014structural,wang2017unravelling}, forming a partially desodiated phase which has a larger inter-layer spacing than pristine NaNiO$_2$. 
	This is due to the reduction in the number of interlayer Na$^+$ cations that screen the Coulombic repulsions between the negatively charged O-Ni-O layers. This phase is very minor, contributing little to the diffraction pattern, and was not accounted for during fitting. 
	It is interesting to note that this peak seems to broaden and gradually merge with the monoclinic [001] peak on application of pressure, disappearing as a peak in its own right by $\sim$30\,GPa (to the nearest 5\,GPa) and resulting simply in peak asymmetry of the [001] peak. 
	The reason for this is likely that the desodiated phase has a high initial compressibility in the direction perpendicular to the layers, and so upon small compression will have an inter-layer spacing approximately equal to the pristine NaNiO$_2$ phase and hence exhibit no signature in the diffraction patterns.
	
	Due to the very low sample volume and the conditions within the DAC, there is a discrete distribution of particle orientations which makes the data unsuitable for Rietveld refinement~\cite{rietveld1969profile}, which would give information on the pressure-dependence of atomic positions. 
	Analysis is therefore limited to Pawley refinement~\cite{pawley1981unit} of lattice parameters from the diffraction data, which is shown in Figure~\ref{LPs_with_pressure} for the 294\,K isothermal compression. 
	Representative Pawley refinements~\cite{pawley1981unit} are shown in Figure~\ref{PawleyRefinementsFig}. 
	In discussing the results of the Pawley refinements, we have focused on the findings of the 294\,K isotherm. 
	The results for the 403\,K isotherm are consistent in terms of unit cell volume, $a/b$ ratio, and intra-/inter-layer distances, as presented in Appendix~2. 
	
	The unit cell volume per formula unit, Figure~\ref{LPs_with_pressure}(a), is lower for the rhombohedral than the monoclinic phase on increasing pressure. 
	This suggests a negative thermal expansion on heating at high-pressure from the monoclinic to rhombohedral structures, which is different to the ambient-pressure behaviour where the rhombohedral is larger than the monoclinic cell~\cite{nagle2024displacive}. 
	The $V(P)$ data is overlaid with a 2nd-order Birch-Murnaghan $P-V$ equation of state~\cite{birch1947finite,angel2000equations} based on parameters fit using PASCAL~\cite{cliffe2012pascal,lertkiattrakul2023pascal}. 
	This fit gives bulk modulus, $B$, of 136.0(4)\,GPa and 138.4(8)\,GPa for the monoclinic and rhombohedral phases respectively. 
	For monoclinic $V(P)$ data, a 3rd-order Birch-Murnaghan equation was needed to fit the data throughout the whole pressure range, as the gradient in bulk modulus, $B'$, deviates from 4. This fit gave values of $B_0=129.4(6)$\,GPa and $B'=4.21(3)$.
	The reference volume was refined in all Birch-Murnaghan fitting. 
	We note that the Birch-Murnaghan equation of state is derived assuming a cubic system~\cite{birch1947finite,angel1996compression,katsura2019simple}, but is widely used for non-cubic systems~\cite{ballaran2003equation,pavese2003equation,zanazzi2007structural,fujimoto2018observation,pavese2005relations}.
	
	The ratio between the $a$ and $b$ lattice parameters would be $\sqrt{3}$ for the rhombohedral phase, but due to the monoclinic distortion $a/b > \sqrt{3}$. 
	On application of pressure, the extent of the deviation of this ratio from $\sqrt{3}$ decreases [Figure~\ref{LPs_with_pressure}(b)], suggesting the cooperatively JT-distorted phase becomes less monoclinic. 
	This may be due to a reduction in the magnitude of the JT distortion reported previously~\cite{nagle2022pressure}, or due to a gradual breakdown in the collinear orbital ordering. 
	
	The inter-layer distance in NaNiO$_2$ is equivalent to $c/3$ for the rhombohedral cell, and $c\cdot\sin{(\beta)}$ for the monoclinic cell. 
	The rhombohedral cell has a larger inter-layer spacing than the monoclinic cell, Figure~\ref{LPs_with_pressure}(c), throughout the pressure range, which is true also at ambient-pressure~\cite{nagle2024displacive}. 
	
	The intra-layer metal-metal interatomic distances are compared in Figure~\ref{LPs_with_pressure}(d). These are $a=b$ for the rhombohedral cell, and $a/\sqrt{3}$ and $b$ for the monoclinic cell. 
	This shows that the rhombohedral intra-layer metal-metal distance is slightly smaller than the average of the two equivalent monoclinic distances throughout the measured pressure range, likely due to the JT distortion pushing oxygen anions towards the metal-metal layer.
	
	\section{Discussion}
	
	\begin{figure*}%[t]
		\includegraphics[scale=2]{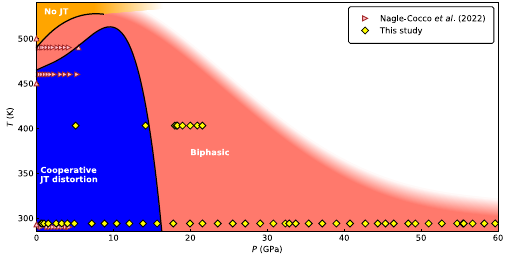}
		\caption{
			A tentative phase diagram of the cooperative JT distortion in NaNiO$_2$, showing the dome-like nature of the cooperatively-JT-distorted region, using data from this work and from Ref.~\cite{nagle2022pressure}. 
			Solid lines indicate known crossovers. 
			The uncoloured region is where the speculative crossover between biphasic and rhombohedral-only may occur. 
			It is not known whether the high-pressure rhombohedral phase will exhibit displacive-like behaviour without local JT distortions, as is the case at ambient pressure and high-$T$~\cite{nagle2024displacive}, or order-disorder behaviour with local JT distortions.
			For the work from the 2022 study, triangles point right if $P$ was increasing or up if $T$ was increasing during the measurement. For the present study, all data points are on increasing pressure. 
		}
		\label{phase_diagram}
	\end{figure*}

	Consistent with previous ambient-pressure and variable-pressure diffraction measurements~\cite{dyer1954alkali,dick1997structure,chappel2000study,sofin2005new,nagle2022pressure,nagle2024displacive}, at the lowest measured pressures at the studied temperatures, NaNiO$_2$ is entirely monoclinic. 
	We have, however, extended the studied pressure range for the JT distortion in NaNiO$_2$. 
	In doing so, we observe a previously unobserved high-pressure rhombohedral phase with the same symmetry as the high-temperature rhombohedral phase. 
	Given that the number of formula units per unit cell and approximate cell shape do not change, we suspect that this rhombohedral phase has the same qualitative average structure and Wyckoff site occupancies as the high-temperature, ambient-pressure unit cell shown in Figure~\ref{NaNiO2-structures.pdf}(c,d). 
	However, it is important to acknowledge that our Pawley analysis is not sensitive to atomic positions or local distortions. 
	
	Assuming an increase in enthalpy on heating, the Clausius-Clapeyron equation states that within the two-phase region, the sign of $dT^\mathrm{JT}_\mathrm{onset}(P)/dP$ will match the sign of $\Delta V = V_\mathrm{rhombohedral}-V_\mathrm{monoclinic}$. 
	We showed in Ref.~\cite{nagle2024displacive} that at ambient pressure $\Delta V > 0$, and in Ref.~\cite{nagle2022pressure} that $dT^\mathrm{JT}_\mathrm{onset}(P)/dP > 0$, consistent with the Clausius-Clapeyron equation. 
	In contrast, Figure~\ref{LPs_with_pressure}(a) shows that at higher pressure, $\Delta V < 0$. 
	We interpret this as indicating that there is a change in sign of $dT^\mathrm{JT}_\mathrm{onset}(P)/dP$ and, consequently, a dome-like shape to the monoclinic region of the phase diagram, albeit with a very broad region of coexistence, as shown tentatively in Figure~\ref{phase_diagram}. 
	The positions where we sketch phase lines are bounded by the pressure-temperature points measured in this work and in Ref.~\cite{nagle2022pressure}. 
	It must be emphasised that there is much that is not known about this phase diagram. It is not known at what pressure the dome is at a maximum (i.e. the pressure for which $dT^\mathrm{JT}_\mathrm{onset}(P)/dP=0$), and so the maximum shown in Figure~\ref{phase_diagram} should be taken as a rough estimate. 
	The line which separates the monoclinic and biphasic regions at high pressure can be tightly constrained by the spacing of measured pressure points, but nothing can be inferred about the position of the line separating the biphasic region and presumed rhombohedral region at high pressure (if it even exists) with the currently available data, and so the range of possible positions of this phase line is shown as a unmarked region in the phase diagram.
	
	%In our previous work~\cite{nagle2022pressure}, we showed by fitting a 2nd-order Birch-Murnaghan equation of state to unit cell volume data in the biphasic regime at 490\,K that the bulk modulus of the rhombohedral phase (113(1)\,GPa) is larger than the monoclinic phase (110(1)\,GPa). 
	%This indicates that the monoclinic phase is more compressible than the rhombohedral phase over the studied pressure range, similar to the Birch-Murnaghan fitting shown in Figure~\ref{LPs_with_pressure} for the ambient-temperature, high-pressure case; although in both cases the difference is small. 
	
	We hypothesise that the inversion of relative volume of the rhombohedral and monoclinic unit cells may be related to the pressure-dependence of the compressibility of the short and long Ni-O bonds within NiO$_6$ octahedra. 
	Our previous work~\cite{nagle2022pressure} showed that the $2\times$long Ni-O bond lengths were far more compressible than the $4\times$short Ni-O bond lengths in the studied pressure range. 
	Studies of other JT-distorted materials such as [(CH$_3$)$_2$NH$_2$][Cu(HCOO)$_3$]~\cite{collings2018disorder}, CuAs$_2$O$_4$~\cite{caslin2016competing}, and CuMnO$_2$~\cite{lawler2021decoupling} show a similar trend at low pressures, but that beyond some pressure the elongated metal-oxygen bonds become significantly less compressible, as seen by a large change in the gradient of bond length against pressure. 
	This may relate to the preference for JT transitions to be first-order, despite a second-order transition being allowed by symmetry in both NaNiO$_2$ and LaMnO$_3$. 
	Such a reduction in compressibility of the long Ni-O bond, if it occurs, would likely reduce the compressibility of the monoclinic phase overall. 
	If the compressibility of the monoclinic phase becomes less than the compressibility of the rhombohedral phase due to a stiffening of the long Ni-O bond, we speculate that further compression of the monoclinic/rhombohedral mixture could eventually lead to the inversion in relative unit cell volume per formula unit seen in Figure~\ref{LPs_with_pressure}(a). 
	However, to test this hypothesis would require variable-pressure diffraction data starting in the biphasic regime (i.e. an isotherm at $\sim$490\,K) which we do not have. 
		
	Our previous work showed that, at ambient pressure, the temperature-induced JT transition is displacive~\cite{nagle2024displacive} with no local JT distortions at high-temperature. 
	However, it is worth noting the result of a previous study which exerted pressure (in the form of uniaxial strain) on SrTiO$_3$~\cite{salmani2020order}. 
	SrTiO$_3$ exhibits a displacive ferroelectric transition at ambient pressure, but, under uniaxial strain, the transition becomes order-disorder with local cation off-centering observed by electron microscopy. 
	Such a scenario for NaNiO$_2$ would suggest a difference in local behaviour between the high-temperature and high-pressure rhombohedral phase. 
	Unfortunately, it is not possible to obtain, by Rietveld refinement~\cite{rietveld1969profile}, the precise atomic positions with pressure from this diffraction data, nor to obtain any insight into the local structure of the high-pressure rhombohedral phase. 
	We therefore do not know whether high-pressure rhombohedral NaNiO$_2$ will show displacive or order-disorder behaviour, and so cannot rule out the possibility that local Jahn--Teller distortions exist in the high-pressure rhombohedral phase. 
	It is left to future work to determine the local structure of high-pressure rhombohedral NaNiO$_2$, either by computational means (such as the \textit{ab initio} molecular dynamics performed previously on layered nickelates~\cite{sicolo2020and,genreith2024jahn,nagle2024displacive}) or by experimental study such as variable-pressure Extended X-ray Absorption Fine Structure (EXAFS)~\cite{rodrigues2023evidence} or total scattering~\cite{herlihy2021recovering}.  
	
	%[We might want to talk about the PbTiO3 stuff. I.e. PbTiO3 shows an order-disorder volume collapse~\cite{rossetti1998phase,pan2022tolerance} like LMO and concurrently $dP/dT<0$~\cite{sani2002pressure} - i.e could volume collapse mean order-disorder? What are implications for high pressure when sign changes, displacive to order disorder in different P regimes?]
	
	\section{Conclusion}
	
	We have studied the effect of pressure (up to 66.58(4)\,GPa) on cooperatively-JT-distorted NaNiO$_2$ and shown that a rhombohedral phase, without a cooperative JT distortion, emerges at high pressure, coexisting with the cooperatively-distorted monoclinic phase throughout the measured pressure range. 
	We infer from the results that there is a dome-like phase diagram for the cooperative JT distortion in NaNiO$_2$. 
	This inference is based on the emergence of the high-pressure rhombohedral phase, in conjunction with an observed decrease in unit cell volume from the monoclinic to the rhombohedral phase and prior evidence for an increasing $T^\mathrm{JT}_\mathrm{onset}(P)$ phase line at low-pressures~\cite{nagle2022pressure,nagle2024displacive}. 
		
	\section*{Appendix 1: pressure calibration}
	\setcounter{subsection}{0}
	
	\renewcommand{\theequation}{A.\arabic{equation}}
	\renewcommand{\thetable}{A.\arabic{table}}
	\renewcommand{\thefigure}{A.\arabic{figure}}
	\setcounter{equation}{0}
	\setcounter{figure}{0}
	\setcounter{table}{0}
	
	\begin{figure}[t]
		\includegraphics[scale=1.0]{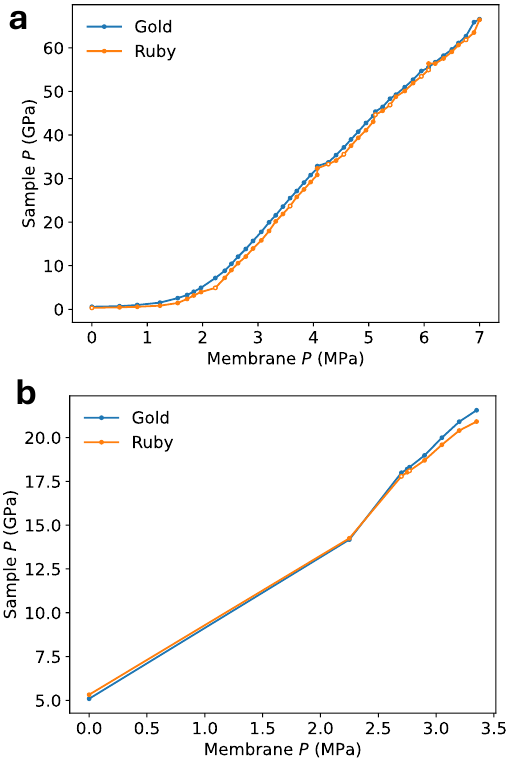}
		\caption{
			Comparison of the results of pressure calibration using the ruby fluorescence and Au diffraction approaches described in Appendix~1, at (a) 294\,K and (b) 403\,K, as a function of pressure in the DAC membrane. 
			Error bars for the pressure obtained from gold are calculated due to the uncertainty in refined lattice parameters, and are excluded as they are smaller than the data points. 
			For the ruby data, open circles denote pressure points where one of the two ruby fluorescence measurements were not performed, either before or after, and so the calculated pressure is based only on the single ruby fluorescence measurement that was performed. 
		}
		\label{calibrant_comparison}
	\end{figure}

	Pressures presented in this paper are those obtained from the Au lattice parameters, not the ruby. 
	Here, we also present ruby calibration to ensure that both calibration methods agree, as shown in Figure~\ref{calibrant_comparison}. 

	\subsection{Pressure calibration using gold}
	
	The calibration of pressure from the Pawley-refined Au lattice parameters followed the thermodynamical equations in Ref.~\cite{mchardy2023creation}. 
	Firstly, at each pressure-temperature point, the unit cell volume, $V$, is calculated from the refined lattice parameter. 
	For a given temperature and unit cell volume, the pressure at a given temperature is obtained from:
	
	\begin{equation}
	P(V,T)=P_\mathrm{BM}(V) + \Delta P_{th}(V,T)
	\end{equation}
	
	Here, $P_\mathrm{BM}$ is obtained \textit{via} the 3rd-order Birch-Murnaghan equation~\cite{birch1947finite,angel2000equations} in which $B_0=169.669$\,GPa and $B'=5.662$ according to Ref.~\cite{mchardy2023creation}. 
	In our calibration, $V_0=67.845$\,\AA{}$^3$, as obtained from Pawley refinement of a diffraction pattern from I15 of Au at ambient pressure and temperature. 
	$\Delta P_{th}(V,T)$ is the thermal component of pressure, given as follows:
	
	\begin{equation}
	\Delta P_{th}(V,T) = \alpha_T(T) B_0 (T) \Delta T(T)
	\end{equation}
	
	Here, $\Delta T(T)=(T-\mathrm{294\,K})$. $\alpha_T(T)$ and $B_0(T)$ are defined as follows:
	
	\begin{equation}
	B_0(T) = \alpha + \Delta T(T) \times \dfrac{dB_0(T)}{dT}
	\end{equation}
	
	\begin{equation}
	\alpha_T(T) = \alpha + \Delta T(T) \times \dfrac{d\alpha(T)}{dT}
	\end{equation}
	
	In these equations, $\alpha=(3.911\times10^{-5})$\,K$^{-1}$, $\dfrac{dB_0(T)}{dT}=0.01017$\,GPa K$^{-1}$, and $\dfrac{d\alpha(T)}{dT}=(1.403\times10^{-8})$\,K$^{-2}$ are empirical parameters taken from Ref.~\cite{mchardy2023creation}.
	
	\subsection{Pressure calibration using ruby fluorescence}
	
	At each pressure-temperature point, the fluorescence spectrum of the ruby was measured. This was typically done before and after measuring a diffraction pattern, though for some pressure-temperature points it was measured only before or after.  
	For each fluorescence spectrum, two pseudo-Voigt peaks were fit, with the following equation used:
	
	\begin{equation}
		\begin{split}
			F(\lambda) = 
			A_1 \exp{\left[ \frac{-(x-x_1)^2}{2\sigma_1^2} \right]} \frac{\Gamma_1^2}{\Gamma_1^2 + (x-x_1)^2}\\
			+ 	A_2 \exp{\left[ \frac{-(x-x_2)^2}{2\sigma_2^2} \right]} \frac{\Gamma_2^2}{\Gamma_2^2 + (x-x_2)^2}
		\end{split}
	\end{equation}
	
	where $F(\lambda)$ is the fluorescence intensity as a function of wavelength, and $A_i$, $\sigma_i$, $x_i$, and $\Gamma_i$ ($i$=1,2) are the intensity, Gaussian width, centre, and Lorentzian width of the peakshapes simulating the $R_1$ and $R_2$ transitions respectively. 
	There were therefore 8 refined parameters for each spectrum, and $x_1$ is taken as $\lambda_0$ for calculation of pressure \textit{via} the following equation~\cite{shen2020toward}:
	
	\begin{equation}
	P = 1.87 \times 10^3 \left( \frac{\Delta\lambda(T)}{\lambda_0(T)} \right) \left[ 1 + 5.63 \frac{\Delta\lambda}{\lambda_0(T)} \right]
	\end{equation}
	
	where $\Delta\lambda(T) = x_1 - \lambda_0(T)$ and $\lambda_0(T)$ is the temperature-dependent, ambient-pressure fluorescence wavelength of a particular ruby. For the ruby used in this experiment, $\lambda_0(300\mathrm{\,K})=694.166$\,nm as determined by fitting the fluorescence spectrum of the ruby at ambient pressure and temperature. To obtain $\lambda_0(T)$, the following relation was used (taken directly from Ref.~\cite{datchi2007optical}):
	
	\begin{equation}
	\lambda_0(T) = \lambda_0(300 K) + (T-300)\times0.00726
	\end{equation}
	
	Thus, from each fluorescence spectrum a pressure was calculated. The reported pressure for the ruby data is obtained by averaging, where possible, the calculated pressure before and after the diffraction measurement.
	
	\subsection{Comparison of pressure calibrants}
	
	Figure~\ref{calibrant_comparison} compares the result of pressure calibration using both the ruby and gold marker. 
	It can be seen that these two calibration methods give consistent results. 

	\section*{Appendix 2: 403\,K isotherm}
	\setcounter{subsection}{0}
	
	Results of Pawley refinement for the 403\,K isotherm are shown in Figure~\ref{403K_LPs}, with those for the 294\,K isotherm discussed in the main text. 
	The results are broadly consistent, except there were no peaks due to a desodiated phase in the 403\,K isotherm. 
	
	In Figure~\ref{T-vol-comparison.pdf} we compare the unit cell volume of the monoclinic and rhombohedral phases over the two isotherms, within the pressure range of the 403\,K isotherm. 
	For the monoclinic phase, at low pressure the high-temperature NaNiO$_2$ has a larger volume, but on compression the unit cell volume is larger at lower temperature, consistent with the fact that monoclinic NaNiO$_2$ gets more compressible on heating as reported previously~\cite{nagle2022pressure}.
	
	\renewcommand{\thefigure}{B.\arabic{figure}}
	\setcounter{figure}{0}
	
	\begin{figure*}[t]
		\includegraphics[scale=1.0]{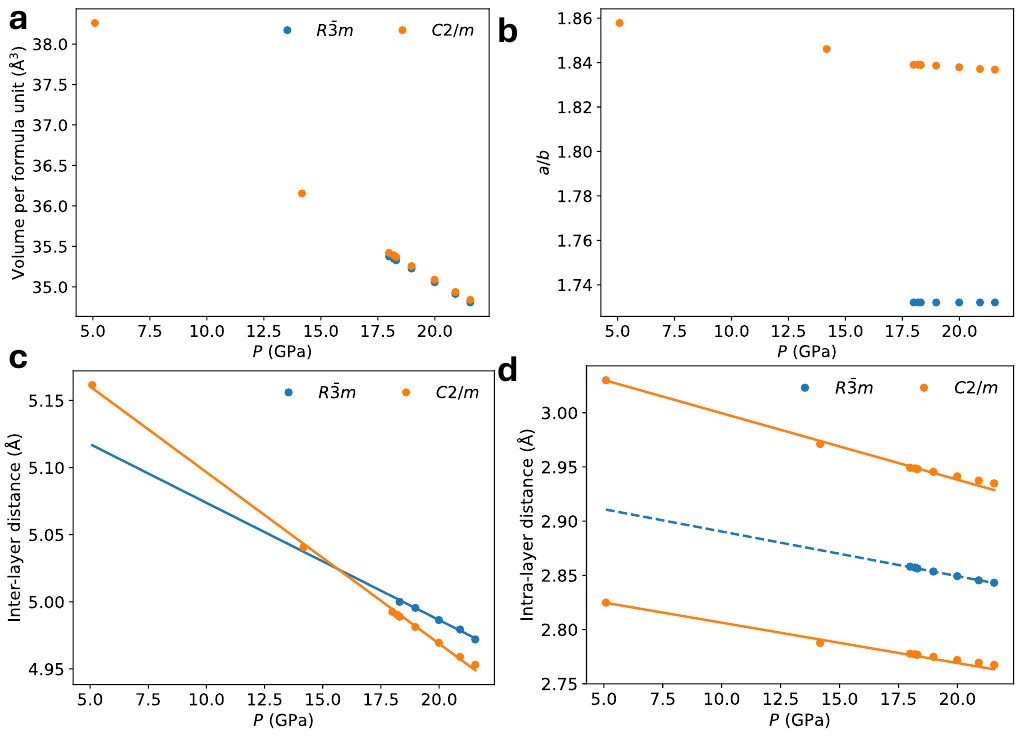}
		\caption{
			The pressure-dependence of the lattice parameters for the 403\,K isotherm. 
			(a) Unit cell volume per formula unit for the rhombohedral (blue) and monoclinic (orange) phases. 
			(b)	Ratio of the monoclinic $a$ and $b$ lattice parameters for the two phases of NaNiO$_2$, showing the reduction of magnitude of monoclinicity of NaNiO$_2$. $a/b=\sqrt{3}$ for the ideal $R\bar{3}m$ structure. 
			(c) inter-layer distances, $c/3$ for the rhombohedral phase and $c\cdot\sin{(\beta)}$ for the monoclinic phase, and (d) intra-layer metal-metal distances, $a$ for the rhombohedral phase, and $a/\sqrt{3}$ and $b$ for the monoclinic phase. 
			In (c) and (d), lines show linear fit to the data (solid lines are the data, dashed line to the average of the monoclinic values), as a visual aid.
		}
		\label{403K_LPs}
	\end{figure*}

	\begin{figure}[t]
	\includegraphics[scale=1.0]{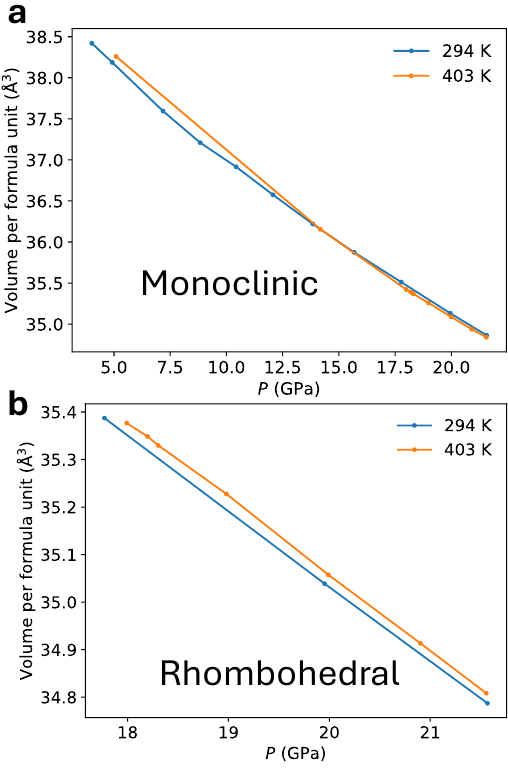}
	\caption{
		A comparison of the unit cell volume per formula unit for the (a) monoclinic and (b) rhomohedral phases over the 294\,K and 403\,K isotherms. 
		Note that errors are smaller than points, and lines join the points and are purely a guide to the eye.
	}
	\label{T-vol-comparison.pdf}
	\end{figure}

	\begin{acknowledgments}
				
		L.A.V.N-C thanks George S. Phillips and Farheen N. Sayed at the University of Cambridge, Craig L. Bull at ISIS Neutron and Muon Source, and Christopher J. Ridley at Oak Ridge National Laboratory for useful discussions. 
		
		The authors thank Diamond Light Source for access to the I15 instrument under beamtime provision CY31718. 
		
		Crystal structure figures were prepared using \textsc{Vesta-3}~\cite{momma2011vesta}. All other figures were prepared using \textsc{Matplotlib}~\cite{Hunter:2007} implemented in \textsc{Python~3}~\cite{hunt2019advanced}.

	\end{acknowledgments}

	\section*{Funding}
	
	We thank Diamond Light Source for beamtime allocation under proposal CY31718. 
	This work was supported by the Faraday Institution (\mbox{FIRG001}, \mbox{FIRG017}, \mbox{FIRG024}, \mbox{FIRG060}). 
	L.A.V.N-C was supported by the UK Engineering and Physical Sciences Research Council (EPSRC) \mbox{(EP/R513180/1)}, along with additional funding from the Cambridge Philosophical Society. 
	J.M.A.S. received support from the EPSRC Cambridge NanoCDT \mbox{(EP/L015978/1}). 
	X.Z. and S.D. were supported by BEIS grant number G115693. 
	S.D. also acknowledges a scholarship to pursue doctoral research from the Cambridge Trust China Scholarship	Council. 
	S.S.S. acknowledges support (G117669) from Department of Science Innovation and Technology, United Kingdom.

	\section*{Data availability}
	
	Data is available in the University of Cambridge repository at doi.org/10.17863/CAM.112461~\cite{doi.org/10.17863/CAM.112461}.
	
	\section*{Corresponding authors}
	
	\begin{itemize}
		\item Liam A. V. Nagle-Cocco (lnc@stanford.edu)
		\item Si\^an E. Dutton (sed33@cam.ac.uk)
	\end{itemize}

	%\section*{Supplementary Information}
	%
	%A document containing Supplementary Information to this article is available. 
	
	\bibliography{references}% Produces the bibliography via BibTeX.

%apsrev4-2.bst 2019-01-14 (MD) hand-edited version of apsrev4-1.bst
%Control: key (0)
%Control: author (8) initials jnrlst
%Control: editor formatted (1) identically to author
%Control: production of article title (0) allowed
%Control: page (0) single
%Control: year (1) truncated
%Control: production of eprint (0) enabled
\begin{thebibliography}{72}%
\makeatletter
\providecommand \@ifxundefined [1]{%
 \@ifx{#1\undefined}
}%
\providecommand \@ifnum [1]{%
 \ifnum #1\expandafter \@firstoftwo
 \else \expandafter \@secondoftwo
 \fi
}%
\providecommand \@ifx [1]{%
 \ifx #1\expandafter \@firstoftwo
 \else \expandafter \@secondoftwo
 \fi
}%
\providecommand \natexlab [1]{#1}%
\providecommand \enquote  [1]{``#1''}%
\providecommand \bibnamefont  [1]{#1}%
\providecommand \bibfnamefont [1]{#1}%
\providecommand \citenamefont [1]{#1}%
\providecommand \href@noop [0]{\@secondoftwo}%
\providecommand \href [0]{\begingroup \@sanitize@url \@href}%
\providecommand \@href[1]{\@@startlink{#1}\@@href}%
\providecommand \@@href[1]{\endgroup#1\@@endlink}%
\providecommand \@sanitize@url [0]{\catcode `\\12\catcode `\$12\catcode
  `\&12\catcode `\#12\catcode `\^12\catcode `\_12\catcode `\%12\relax}%
\providecommand \@@startlink[1]{}%
\providecommand \@@endlink[0]{}%
\providecommand \url  [0]{\begingroup\@sanitize@url \@url }%
\providecommand \@url [1]{\endgroup\@href {#1}{\urlprefix }}%
\providecommand \urlprefix  [0]{URL }%
\providecommand \Eprint [0]{\href }%
\providecommand \doibase [0]{https://doi.org/}%
\providecommand \selectlanguage [0]{\@gobble}%
\providecommand \bibinfo  [0]{\@secondoftwo}%
\providecommand \bibfield  [0]{\@secondoftwo}%
\providecommand \translation [1]{[#1]}%
\providecommand \BibitemOpen [0]{}%
\providecommand \bibitemStop [0]{}%
\providecommand \bibitemNoStop [0]{.\EOS\space}%
\providecommand \EOS [0]{\spacefactor3000\relax}%
\providecommand \BibitemShut  [1]{\csname bibitem#1\endcsname}%
\let\auto@bib@innerbib\@empty
%</preamble>
\bibitem [{\citenamefont {Jahn}\ and\ \citenamefont
  {Teller}(1937)}]{jahn1937stability}%
  \BibitemOpen
  \bibfield  {author} {\bibinfo {author} {\bibfnamefont {H.~A.}\ \bibnamefont
  {Jahn}}\ and\ \bibinfo {author} {\bibfnamefont {E.}~\bibnamefont {Teller}},\
  }\bibfield  {title} {\bibinfo {title} {Stability of polyatomic molecules in
  degenerate electronic states-{I}—orbital degeneracy},\ }\href
  {https://royalsocietypublishing.org/doi/abs/10.1098/rspa.1937.0142}
  {\bibfield  {journal} {\bibinfo  {journal} {Proceedings of the Royal Society
  of London. Series A-Mathematical and Physical Sciences}\ }\textbf {\bibinfo
  {volume} {161}},\ \bibinfo {pages} {220} (\bibinfo {year}
  {1937})}\BibitemShut {NoStop}%
\bibitem [{\citenamefont {Van~Vleck}(1939)}]{van1939jahn}%
  \BibitemOpen
  \bibfield  {author} {\bibinfo {author} {\bibfnamefont {J.~H.}\ \bibnamefont
  {Van~Vleck}},\ }\bibfield  {title} {\bibinfo {title} {The {Jahn--T}eller
  effect and crystalline {S}tark splitting for clusters of the form {XY}$_6$},\
  }\href {https://doi.org/10.1063/1.1750327} {\bibfield  {journal} {\bibinfo
  {journal} {The Journal of Chemical Physics}\ }\textbf {\bibinfo {volume}
  {7}},\ \bibinfo {pages} {72} (\bibinfo {year} {1939})}\BibitemShut {NoStop}%
\bibitem [{\citenamefont {Nagle-Cocco}\ and\ \citenamefont
  {Dutton}(2024)}]{nagle2024van}%
  \BibitemOpen
  \bibfield  {author} {\bibinfo {author} {\bibfnamefont {L.~A.~V.}\
  \bibnamefont {Nagle-Cocco}}\ and\ \bibinfo {author} {\bibfnamefont {S.~E.}\
  \bibnamefont {Dutton}},\ }\bibfield  {title} {\bibinfo {title} {Van {Vleck}
  analysis of angularly distorted octahedra using {VanVleckCalculator}},\
  }\href {https://journals.iucr.org/j/issues/2024/01/00/oc5030/oc5030.pdf}
  {\bibfield  {journal} {\bibinfo  {journal} {IUCr Journal of Applied
  Crystallography}\ }\textbf {\bibinfo {volume} {57}},\ \bibinfo {pages} {20}
  (\bibinfo {year} {2024})}\BibitemShut {NoStop}%
\bibitem [{\citenamefont {Li}\ \emph {et~al.}(2016)\citenamefont {Li},
  \citenamefont {Wang}, \citenamefont {Wu}, \citenamefont {Liu}, \citenamefont
  {Bo},\ and\ \citenamefont {Ceder}}]{li2016jahn}%
  \BibitemOpen
  \bibfield  {author} {\bibinfo {author} {\bibfnamefont {X.}~\bibnamefont
  {Li}}, \bibinfo {author} {\bibfnamefont {Y.}~\bibnamefont {Wang}}, \bibinfo
  {author} {\bibfnamefont {D.}~\bibnamefont {Wu}}, \bibinfo {author}
  {\bibfnamefont {L.}~\bibnamefont {Liu}}, \bibinfo {author} {\bibfnamefont
  {S.-H.}\ \bibnamefont {Bo}},\ and\ \bibinfo {author} {\bibfnamefont
  {G.}~\bibnamefont {Ceder}},\ }\bibfield  {title} {\bibinfo {title}
  {Jahn--{T}eller assisted {N}a diffusion for high performance {N}a ion
  batteries},\ }\href
  {https://pubs.acs.org/doi/full/10.1021/acs.chemmater.6b02440} {\bibfield
  {journal} {\bibinfo  {journal} {Chemistry of Materials}\ }\textbf {\bibinfo
  {volume} {28}},\ \bibinfo {pages} {6575} (\bibinfo {year}
  {2016})}\BibitemShut {NoStop}%
\bibitem [{\citenamefont {Fil}\ \emph {et~al.}(1992)\citenamefont {Fil},
  \citenamefont {Tokar}, \citenamefont {Shelankov},\ and\ \citenamefont
  {Weber}}]{fil1992lattice}%
  \BibitemOpen
  \bibfield  {author} {\bibinfo {author} {\bibfnamefont {D.~V.}\ \bibnamefont
  {Fil}}, \bibinfo {author} {\bibfnamefont {O.~I.}\ \bibnamefont {Tokar}},
  \bibinfo {author} {\bibfnamefont {A.~L.}\ \bibnamefont {Shelankov}},\ and\
  \bibinfo {author} {\bibfnamefont {W.}~\bibnamefont {Weber}},\ }\bibfield
  {title} {\bibinfo {title} {Lattice-mediated interaction of {Cu}$^{2+}$
  {Jahn--Teller} ions in insulating cuprates},\ }\href
  {https://journals.aps.org/prb/abstract/10.1103/PhysRevB.45.5633} {\bibfield
  {journal} {\bibinfo  {journal} {Physical Review B}\ }\textbf {\bibinfo
  {volume} {45}},\ \bibinfo {pages} {5633} (\bibinfo {year}
  {1992})}\BibitemShut {NoStop}%
\bibitem [{\citenamefont {Keller}\ \emph {et~al.}(2008)\citenamefont {Keller},
  \citenamefont {Bussmann-Holder},\ and\ \citenamefont
  {M{\"u}ller}}]{keller2008jahn}%
  \BibitemOpen
  \bibfield  {author} {\bibinfo {author} {\bibfnamefont {H.}~\bibnamefont
  {Keller}}, \bibinfo {author} {\bibfnamefont {A.}~\bibnamefont
  {Bussmann-Holder}},\ and\ \bibinfo {author} {\bibfnamefont {K.~A.}\
  \bibnamefont {M{\"u}ller}},\ }\bibfield  {title} {\bibinfo {title}
  {{Jahn--Teller} physics and high-${T}_c$ superconductivity},\ }\href
  {https://www.sciencedirect.com/science/article/pii/S1369702108701780}
  {\bibfield  {journal} {\bibinfo  {journal} {Materials Today}\ }\textbf
  {\bibinfo {volume} {11}},\ \bibinfo {pages} {38} (\bibinfo {year}
  {2008})}\BibitemShut {NoStop}%
\bibitem [{\citenamefont {Merten}\ \emph {et~al.}(2019)\citenamefont {Merten},
  \citenamefont {Shapoval}, \citenamefont {Damaschke}, \citenamefont {Samwer},\
  and\ \citenamefont {Moshnyaga}}]{merten2019magnetic}%
  \BibitemOpen
  \bibfield  {author} {\bibinfo {author} {\bibfnamefont {S.}~\bibnamefont
  {Merten}}, \bibinfo {author} {\bibfnamefont {O.}~\bibnamefont {Shapoval}},
  \bibinfo {author} {\bibfnamefont {B.}~\bibnamefont {Damaschke}}, \bibinfo
  {author} {\bibfnamefont {K.}~\bibnamefont {Samwer}},\ and\ \bibinfo {author}
  {\bibfnamefont {V.}~\bibnamefont {Moshnyaga}},\ }\bibfield  {title} {\bibinfo
  {title} {Magnetic-field-induced suppression of {Jahn--Teller} phonon bands in
  ({L}a$_{0.6}${P}r$_{0.4}$)$_{0.7}${C}a$_{0.3}${M}n{O}$_3$: the mechanism of
  colossal magnetoresistance shown by {Raman} spectroscopy},\ }\href
  {https://www.nature.com/articles/s41598-019-39597-1} {\bibfield  {journal}
  {\bibinfo  {journal} {Scientific Reports}\ }\textbf {\bibinfo {volume} {9}},\
  \bibinfo {pages} {2387} (\bibinfo {year} {2019})}\BibitemShut {NoStop}%
\bibitem [{\citenamefont {{\AA}sbrink}\ \emph {et~al.}(1999)\citenamefont
  {{\AA}sbrink}, \citenamefont {Wa{\'s}kowska}, \citenamefont {Gerward},
  \citenamefont {Olsen},\ and\ \citenamefont {Talik}}]{aasbrink1999high}%
  \BibitemOpen
  \bibfield  {author} {\bibinfo {author} {\bibfnamefont {S.}~\bibnamefont
  {{\AA}sbrink}}, \bibinfo {author} {\bibfnamefont {A.}~\bibnamefont
  {Wa{\'s}kowska}}, \bibinfo {author} {\bibfnamefont {L.}~\bibnamefont
  {Gerward}}, \bibinfo {author} {\bibfnamefont {J.~S.}\ \bibnamefont {Olsen}},\
  and\ \bibinfo {author} {\bibfnamefont {E.}~\bibnamefont {Talik}},\ }\bibfield
   {title} {\bibinfo {title} {High-pressure phase transition and properties of
  spinel {Z}n{M}n$_2${O}$_4$},\ }\href
  {https://journals.aps.org/prb/abstract/10.1103/PhysRevB.60.12651} {\bibfield
  {journal} {\bibinfo  {journal} {Phys. Rev. B}\ }\textbf {\bibinfo {volume}
  {60}},\ \bibinfo {pages} {12651} (\bibinfo {year} {1999})}\BibitemShut
  {NoStop}%
\bibitem [{\citenamefont {Loa}\ \emph {et~al.}(2001)\citenamefont {Loa},
  \citenamefont {Adler}, \citenamefont {Grzechnik}, \citenamefont {Syassen},
  \citenamefont {Schwarz}, \citenamefont {Hanfland}, \citenamefont {Rozenberg},
  \citenamefont {Gorodetsky},\ and\ \citenamefont
  {Pasternak}}]{loa2001pressure}%
  \BibitemOpen
  \bibfield  {author} {\bibinfo {author} {\bibfnamefont {I.}~\bibnamefont
  {Loa}}, \bibinfo {author} {\bibfnamefont {P.}~\bibnamefont {Adler}}, \bibinfo
  {author} {\bibfnamefont {A.}~\bibnamefont {Grzechnik}}, \bibinfo {author}
  {\bibfnamefont {K.}~\bibnamefont {Syassen}}, \bibinfo {author} {\bibfnamefont
  {U.}~\bibnamefont {Schwarz}}, \bibinfo {author} {\bibfnamefont
  {M.}~\bibnamefont {Hanfland}}, \bibinfo {author} {\bibfnamefont {G.~K.}\
  \bibnamefont {Rozenberg}}, \bibinfo {author} {\bibfnamefont {P.}~\bibnamefont
  {Gorodetsky}},\ and\ \bibinfo {author} {\bibfnamefont {M.~P.}\ \bibnamefont
  {Pasternak}},\ }\bibfield  {title} {\bibinfo {title} {Pressure-induced
  quenching of the {J}ahn--{T}eller distortion and insulator-to-metal
  transition in {LaMnO}$_3$},\ }\href
  {https://journals.aps.org/prl/abstract/10.1103/PhysRevLett.87.125501}
  {\bibfield  {journal} {\bibinfo  {journal} {Phys. Rev. Lett.}\ }\textbf
  {\bibinfo {volume} {87}},\ \bibinfo {pages} {125501} (\bibinfo {year}
  {2001})}\BibitemShut {NoStop}%
\bibitem [{\citenamefont {Waskowska}\ \emph {et~al.}(2001)\citenamefont
  {Waskowska}, \citenamefont {Gerward}, \citenamefont {Staun~Olsen},
  \citenamefont {Steenstrup},\ and\ \citenamefont
  {Talik}}]{waskowska2001cumn2o4}%
  \BibitemOpen
  \bibfield  {author} {\bibinfo {author} {\bibfnamefont {A.}~\bibnamefont
  {Waskowska}}, \bibinfo {author} {\bibfnamefont {L.}~\bibnamefont {Gerward}},
  \bibinfo {author} {\bibfnamefont {J.}~\bibnamefont {Staun~Olsen}}, \bibinfo
  {author} {\bibfnamefont {S.}~\bibnamefont {Steenstrup}},\ and\ \bibinfo
  {author} {\bibfnamefont {E.}~\bibnamefont {Talik}},\ }\bibfield  {title}
  {\bibinfo {title} {{CuMn}$_2${O}$_4$: properties and the high-pressure
  induced {J}ahn--{T}eller phase transition},\ }\href
  {https://iopscience.iop.org/article/10.1088/0953-8984/13/11/311/pdf}
  {\bibfield  {journal} {\bibinfo  {journal} {Journal of Physics: Condensed
  Matter}\ }\textbf {\bibinfo {volume} {13}},\ \bibinfo {pages} {2549}
  (\bibinfo {year} {2001})}\BibitemShut {NoStop}%
\bibitem [{\citenamefont {Zhou}\ \emph {et~al.}(2008)\citenamefont {Zhou},
  \citenamefont {Uwatoko}, \citenamefont {Matsubayashi},\ and\ \citenamefont
  {Goodenough}}]{zhou2008breakdown}%
  \BibitemOpen
  \bibfield  {author} {\bibinfo {author} {\bibfnamefont {J.-S.}\ \bibnamefont
  {Zhou}}, \bibinfo {author} {\bibfnamefont {Y.}~\bibnamefont {Uwatoko}},
  \bibinfo {author} {\bibfnamefont {K.}~\bibnamefont {Matsubayashi}},\ and\
  \bibinfo {author} {\bibfnamefont {J.~B.}\ \bibnamefont {Goodenough}},\
  }\bibfield  {title} {\bibinfo {title} {Breakdown of magnetic order in {M}ott
  insulators with frustrated superexchange interaction},\ }\href
  {https://journals.aps.org/prb/abstract/10.1103/PhysRevB.78.220402} {\bibfield
   {journal} {\bibinfo  {journal} {Phys. Rev. B}\ }\textbf {\bibinfo {volume}
  {78}},\ \bibinfo {pages} {220402} (\bibinfo {year} {2008})}\BibitemShut
  {NoStop}%
\bibitem [{\citenamefont {Zhou}\ \emph {et~al.}(2011)\citenamefont {Zhou},
  \citenamefont {Alonso}, \citenamefont {Han}, \citenamefont
  {Fern{\'a}ndez-D{\'\i}az}, \citenamefont {Cheng},\ and\ \citenamefont
  {Goodenough}}]{zhou2011jahn}%
  \BibitemOpen
  \bibfield  {author} {\bibinfo {author} {\bibfnamefont {J.-S.}\ \bibnamefont
  {Zhou}}, \bibinfo {author} {\bibfnamefont {J.~A.}\ \bibnamefont {Alonso}},
  \bibinfo {author} {\bibfnamefont {J.~T.}\ \bibnamefont {Han}}, \bibinfo
  {author} {\bibfnamefont {M.~T.}\ \bibnamefont {Fern{\'a}ndez-D{\'\i}az}},
  \bibinfo {author} {\bibfnamefont {J.-G.}\ \bibnamefont {Cheng}},\ and\
  \bibinfo {author} {\bibfnamefont {J.~B.}\ \bibnamefont {Goodenough}},\
  }\bibfield  {title} {\bibinfo {title} {Jahn--{T}eller distortion in
  perovskite {KCuF}$_3$ under high pressure},\ }\href
  {https://www.sciencedirect.com/science/article/pii/S0022113911002491}
  {\bibfield  {journal} {\bibinfo  {journal} {Journal of Fluorine Chemistry}\
  }\textbf {\bibinfo {volume} {132}},\ \bibinfo {pages} {1117} (\bibinfo {year}
  {2011})}\BibitemShut {NoStop}%
\bibitem [{\citenamefont {Aguado}\ \emph {et~al.}(2012)\citenamefont {Aguado},
  \citenamefont {Rodr{\'\i}guez}, \citenamefont {Valiente}, \citenamefont
  {Iti{\`e}},\ and\ \citenamefont {Hanfland}}]{aguado2012pressure}%
  \BibitemOpen
  \bibfield  {author} {\bibinfo {author} {\bibfnamefont {F.}~\bibnamefont
  {Aguado}}, \bibinfo {author} {\bibfnamefont {F.}~\bibnamefont
  {Rodr{\'\i}guez}}, \bibinfo {author} {\bibfnamefont {R.}~\bibnamefont
  {Valiente}}, \bibinfo {author} {\bibfnamefont {J.-P.}\ \bibnamefont
  {Iti{\`e}}},\ and\ \bibinfo {author} {\bibfnamefont {M.}~\bibnamefont
  {Hanfland}},\ }\bibfield  {title} {\bibinfo {title} {Pressure effects on
  {J}ahn--{T}eller distortion in perovskites: The roles of local and bulk
  compressibilities},\ }\href {https://doi.org/10.1103/PhysRevB.85.100101}
  {\bibfield  {journal} {\bibinfo  {journal} {Physical Review B}\ }\textbf
  {\bibinfo {volume} {85}},\ \bibinfo {pages} {100101} (\bibinfo {year}
  {2012})}\BibitemShut {NoStop}%
\bibitem [{\citenamefont {Mota}\ \emph {et~al.}(2014)\citenamefont {Mota},
  \citenamefont {Almeida}, \citenamefont {Rodrigues}, \citenamefont {Costa},
  \citenamefont {Tavares}, \citenamefont {Bouvier}, \citenamefont {Guennou},
  \citenamefont {Kreisel},\ and\ \citenamefont {Moreira}}]{mota2014dynamic}%
  \BibitemOpen
  \bibfield  {author} {\bibinfo {author} {\bibfnamefont {D.~A.}\ \bibnamefont
  {Mota}}, \bibinfo {author} {\bibfnamefont {A.}~\bibnamefont {Almeida}},
  \bibinfo {author} {\bibfnamefont {V.~H.}\ \bibnamefont {Rodrigues}}, \bibinfo
  {author} {\bibfnamefont {M.~M.~R.}\ \bibnamefont {Costa}}, \bibinfo {author}
  {\bibfnamefont {P.}~\bibnamefont {Tavares}}, \bibinfo {author} {\bibfnamefont
  {P.}~\bibnamefont {Bouvier}}, \bibinfo {author} {\bibfnamefont
  {M.}~\bibnamefont {Guennou}}, \bibinfo {author} {\bibfnamefont
  {J.}~\bibnamefont {Kreisel}},\ and\ \bibinfo {author} {\bibfnamefont {J.~A.}\
  \bibnamefont {Moreira}},\ }\bibfield  {title} {\bibinfo {title} {Dynamic and
  structural properties of orthorhombic rare-earth manganites under high
  pressure},\ }\href {https://doi.org/10.1103/PhysRevB.90.054104} {\bibfield
  {journal} {\bibinfo  {journal} {Physical Review B}\ }\textbf {\bibinfo
  {volume} {90}},\ \bibinfo {pages} {054104} (\bibinfo {year}
  {2014})}\BibitemShut {NoStop}%
\bibitem [{\citenamefont {Caslin}\ \emph {et~al.}(2016)\citenamefont {Caslin},
  \citenamefont {Kremer}, \citenamefont {Razavi}, \citenamefont {Hanfland},
  \citenamefont {Syassen}, \citenamefont {Gordon},\ and\ \citenamefont
  {Whangbo}}]{caslin2016competing}%
  \BibitemOpen
  \bibfield  {author} {\bibinfo {author} {\bibfnamefont {K.}~\bibnamefont
  {Caslin}}, \bibinfo {author} {\bibfnamefont {R.~K.}\ \bibnamefont {Kremer}},
  \bibinfo {author} {\bibfnamefont {F.~S.}\ \bibnamefont {Razavi}}, \bibinfo
  {author} {\bibfnamefont {M.}~\bibnamefont {Hanfland}}, \bibinfo {author}
  {\bibfnamefont {K.}~\bibnamefont {Syassen}}, \bibinfo {author} {\bibfnamefont
  {E.~E.}\ \bibnamefont {Gordon}},\ and\ \bibinfo {author} {\bibfnamefont
  {M.-H.}\ \bibnamefont {Whangbo}},\ }\bibfield  {title} {\bibinfo {title}
  {Competing {J}ahn--{T}eller distortions and hydrostatic pressure effects in
  the quasi-one-dimensional quantum ferromagnet {C}u{A}s$_2${O}$_4$},\ }\href
  {https://journals.aps.org/prb/abstract/10.1103/PhysRevB.93.022301} {\bibfield
   {journal} {\bibinfo  {journal} {Phys. Rev. B}\ }\textbf {\bibinfo {volume}
  {93}},\ \bibinfo {pages} {022301} (\bibinfo {year} {2016})}\BibitemShut
  {NoStop}%
\bibitem [{\citenamefont {Zhao}\ \emph {et~al.}(2016)\citenamefont {Zhao},
  \citenamefont {Yang}, \citenamefont {Li}, \citenamefont {Li}, \citenamefont
  {Tang}, \citenamefont {Li}, \citenamefont {Zhu}, \citenamefont {Zhu},\ and\
  \citenamefont {Wang}}]{zhao2016pressure}%
  \BibitemOpen
  \bibfield  {author} {\bibinfo {author} {\bibfnamefont {Y.}~\bibnamefont
  {Zhao}}, \bibinfo {author} {\bibfnamefont {W.}~\bibnamefont {Yang}}, \bibinfo
  {author} {\bibfnamefont {N.}~\bibnamefont {Li}}, \bibinfo {author}
  {\bibfnamefont {Y.}~\bibnamefont {Li}}, \bibinfo {author} {\bibfnamefont
  {R.}~\bibnamefont {Tang}}, \bibinfo {author} {\bibfnamefont {H.}~\bibnamefont
  {Li}}, \bibinfo {author} {\bibfnamefont {H.}~\bibnamefont {Zhu}}, \bibinfo
  {author} {\bibfnamefont {P.}~\bibnamefont {Zhu}},\ and\ \bibinfo {author}
  {\bibfnamefont {X.}~\bibnamefont {Wang}},\ }\bibfield  {title} {\bibinfo
  {title} {Pressure-enhanced insulating state and trigonal distortion
  relaxation in geometrically frustrated pyrochlore {E}u$_2${S}n$_2${O}$_7$},\
  }\href {https://doi.org/10.1021/acs.jpcc.6b02246} {\bibfield  {journal}
  {\bibinfo  {journal} {The Journal of Physical Chemistry C}\ }\textbf
  {\bibinfo {volume} {120}},\ \bibinfo {pages} {9436} (\bibinfo {year}
  {2016})}\BibitemShut {NoStop}%
\bibitem [{\citenamefont {Collings}\ \emph {et~al.}(2018)\citenamefont
  {Collings}, \citenamefont {Bykov}, \citenamefont {Bykova}, \citenamefont
  {Hanfland}, \citenamefont {van Smaalen}, \citenamefont {Dubrovinsky},\ and\
  \citenamefont {Dubrovinskaia}}]{collings2018disorder}%
  \BibitemOpen
  \bibfield  {author} {\bibinfo {author} {\bibfnamefont {I.~E.}\ \bibnamefont
  {Collings}}, \bibinfo {author} {\bibfnamefont {M.}~\bibnamefont {Bykov}},
  \bibinfo {author} {\bibfnamefont {E.}~\bibnamefont {Bykova}}, \bibinfo
  {author} {\bibfnamefont {M.}~\bibnamefont {Hanfland}}, \bibinfo {author}
  {\bibfnamefont {S.}~\bibnamefont {van Smaalen}}, \bibinfo {author}
  {\bibfnamefont {L.}~\bibnamefont {Dubrovinsky}},\ and\ \bibinfo {author}
  {\bibfnamefont {N.}~\bibnamefont {Dubrovinskaia}},\ }\bibfield  {title}
  {\bibinfo {title} {Disorder--order transitions in the perovskite
  metal--organic frameworks [({CH}$_3$)$_2${NH}$_2$][{M}({HCOO})$_3$] at high
  pressure},\ }\href
  {https://pubs.rsc.org/en/content/articlehtml/2018/ce/c8ce00617b} {\bibfield
  {journal} {\bibinfo  {journal} {CrystEngComm}\ }\textbf {\bibinfo {volume}
  {20}},\ \bibinfo {pages} {3512} (\bibinfo {year} {2018})}\BibitemShut
  {NoStop}%
\bibitem [{\citenamefont {Bostr{\"o}m}\ \emph {et~al.}(2019)\citenamefont
  {Bostr{\"o}m}, \citenamefont {Collings}, \citenamefont {Cairns},
  \citenamefont {Romao},\ and\ \citenamefont {Goodwin}}]{bostrom2019high}%
  \BibitemOpen
  \bibfield  {author} {\bibinfo {author} {\bibfnamefont {H.~L.~B.}\
  \bibnamefont {Bostr{\"o}m}}, \bibinfo {author} {\bibfnamefont {I.~E.}\
  \bibnamefont {Collings}}, \bibinfo {author} {\bibfnamefont {A.~B.}\
  \bibnamefont {Cairns}}, \bibinfo {author} {\bibfnamefont {C.~P.}\
  \bibnamefont {Romao}},\ and\ \bibinfo {author} {\bibfnamefont {A.~L.}\
  \bibnamefont {Goodwin}},\ }\bibfield  {title} {\bibinfo {title}
  {High-pressure behaviour of prussian blue analogues: interplay of hydration,
  {Jahn--Teller} distortions and vacancies},\ }\href
  {https://pubs.rsc.org/en/content/articlehtml/2018/dt/c8dt04463e} {\bibfield
  {journal} {\bibinfo  {journal} {Dalton Transactions}\ }\textbf {\bibinfo
  {volume} {48}},\ \bibinfo {pages} {1647} (\bibinfo {year}
  {2019})}\BibitemShut {NoStop}%
\bibitem [{\citenamefont {Li}\ \emph {et~al.}(2020)\citenamefont {Li},
  \citenamefont {Liu}, \citenamefont {Dong}, \citenamefont {Li}, \citenamefont
  {Dong}, \citenamefont {Lin}, \citenamefont {Liu}, \citenamefont {Wang},
  \citenamefont {Shen}, \citenamefont {Li},\ and\ \citenamefont
  {Liu}}]{li2020size}%
  \BibitemOpen
  \bibfield  {author} {\bibinfo {author} {\bibfnamefont {J.}~\bibnamefont
  {Li}}, \bibinfo {author} {\bibfnamefont {B.}~\bibnamefont {Liu}}, \bibinfo
  {author} {\bibfnamefont {J.}~\bibnamefont {Dong}}, \bibinfo {author}
  {\bibfnamefont {C.}~\bibnamefont {Li}}, \bibinfo {author} {\bibfnamefont
  {Q.}~\bibnamefont {Dong}}, \bibinfo {author} {\bibfnamefont {T.}~\bibnamefont
  {Lin}}, \bibinfo {author} {\bibfnamefont {R.}~\bibnamefont {Liu}}, \bibinfo
  {author} {\bibfnamefont {P.}~\bibnamefont {Wang}}, \bibinfo {author}
  {\bibfnamefont {P.}~\bibnamefont {Shen}}, \bibinfo {author} {\bibfnamefont
  {Q.}~\bibnamefont {Li}},\ and\ \bibinfo {author} {\bibfnamefont
  {B.}~\bibnamefont {Liu}},\ }\bibfield  {title} {\bibinfo {title} {Size and
  morphology effects on the high pressure behaviors of {M}n$_3${O}$_4$
  nanorods},\ }\href
  {https://pubs.rsc.org/en/content/articlehtml/2020/na/d0na00610f} {\bibfield
  {journal} {\bibinfo  {journal} {Nanoscale Advances}\ }\textbf {\bibinfo
  {volume} {2}},\ \bibinfo {pages} {5841} (\bibinfo {year} {2020})}\BibitemShut
  {NoStop}%
\bibitem [{\citenamefont {Bhadram}\ \emph {et~al.}(2021)\citenamefont
  {Bhadram}, \citenamefont {Joseph}, \citenamefont {Delmonte}, \citenamefont
  {Gilioli}, \citenamefont {Baptiste}, \citenamefont {Le~Godec}, \citenamefont
  {Lobo},\ and\ \citenamefont {Gauzzi}}]{bhadram2021pressure}%
  \BibitemOpen
  \bibfield  {author} {\bibinfo {author} {\bibfnamefont {V.~S.}\ \bibnamefont
  {Bhadram}}, \bibinfo {author} {\bibfnamefont {B.}~\bibnamefont {Joseph}},
  \bibinfo {author} {\bibfnamefont {D.}~\bibnamefont {Delmonte}}, \bibinfo
  {author} {\bibfnamefont {E.}~\bibnamefont {Gilioli}}, \bibinfo {author}
  {\bibfnamefont {B.}~\bibnamefont {Baptiste}}, \bibinfo {author}
  {\bibfnamefont {Y.}~\bibnamefont {Le~Godec}}, \bibinfo {author}
  {\bibfnamefont {R.~P. S.~M.}\ \bibnamefont {Lobo}},\ and\ \bibinfo {author}
  {\bibfnamefont {A.}~\bibnamefont {Gauzzi}},\ }\bibfield  {title} {\bibinfo
  {title} {Pressure-induced structural phase transition and suppression of
  {J}ahn--{T}eller distortion in the quadruple perovskite structure},\ }\href
  {https://journals.aps.org/prmaterials/abstract/10.1103/PhysRevMaterials.5.104411}
  {\bibfield  {journal} {\bibinfo  {journal} {Physical Review Materials}\
  }\textbf {\bibinfo {volume} {5}},\ \bibinfo {pages} {104411} (\bibinfo {year}
  {2021})}\BibitemShut {NoStop}%
\bibitem [{\citenamefont {Lawler}\ \emph {et~al.}(2021)\citenamefont {Lawler},
  \citenamefont {Smith}, \citenamefont {Evans}, \citenamefont {Dos~Santos},
  \citenamefont {Molaison}, \citenamefont {Bos}, \citenamefont {Mutka},
  \citenamefont {Henry}, \citenamefont {Argyriou}, \citenamefont {Salamat},\
  and\ \citenamefont {Kimber}}]{lawler2021decoupling}%
  \BibitemOpen
  \bibfield  {author} {\bibinfo {author} {\bibfnamefont {K.~V.}\ \bibnamefont
  {Lawler}}, \bibinfo {author} {\bibfnamefont {D.}~\bibnamefont {Smith}},
  \bibinfo {author} {\bibfnamefont {S.~R.}\ \bibnamefont {Evans}}, \bibinfo
  {author} {\bibfnamefont {A.~M.}\ \bibnamefont {Dos~Santos}}, \bibinfo
  {author} {\bibfnamefont {J.~J.}\ \bibnamefont {Molaison}}, \bibinfo {author}
  {\bibfnamefont {J.-W.~G.}\ \bibnamefont {Bos}}, \bibinfo {author}
  {\bibfnamefont {H.}~\bibnamefont {Mutka}}, \bibinfo {author} {\bibfnamefont
  {P.~F.}\ \bibnamefont {Henry}}, \bibinfo {author} {\bibfnamefont {D.~N.}\
  \bibnamefont {Argyriou}}, \bibinfo {author} {\bibfnamefont {A.}~\bibnamefont
  {Salamat}},\ and\ \bibinfo {author} {\bibfnamefont {S.~A.~J.}\ \bibnamefont
  {Kimber}},\ }\bibfield  {title} {\bibinfo {title} {Decoupling lattice and
  magnetic instabilities in frustrated {C}u{M}n{O}$_2$},\ }\href
  {https://pubs.acs.org/doi/full/10.1021/acs.inorgchem.1c00435} {\bibfield
  {journal} {\bibinfo  {journal} {Inorganic Chemistry}\ }\textbf {\bibinfo
  {volume} {60}},\ \bibinfo {pages} {6004} (\bibinfo {year}
  {2021})}\BibitemShut {NoStop}%
\bibitem [{\citenamefont {Scatena}\ \emph {et~al.}(2021)\citenamefont
  {Scatena}, \citenamefont {Andrzejewski}, \citenamefont {Johnson},\ and\
  \citenamefont {Macchi}}]{scatena2021pressure}%
  \BibitemOpen
  \bibfield  {author} {\bibinfo {author} {\bibfnamefont {R.}~\bibnamefont
  {Scatena}}, \bibinfo {author} {\bibfnamefont {M.}~\bibnamefont
  {Andrzejewski}}, \bibinfo {author} {\bibfnamefont {R.~D.}\ \bibnamefont
  {Johnson}},\ and\ \bibinfo {author} {\bibfnamefont {P.}~\bibnamefont
  {Macchi}},\ }\bibfield  {title} {\bibinfo {title} {Pressure-induced
  {Jahn--Teller} switch in the homoleptic hybrid perovskite
  [({CH}$_3$)$_2${NH}$_2$]{C}u({HCOO})$_3$: orbital reordering by
  unconventional degrees of freedom},\ }\href
  {https://doi.org/10.1039/D1TC01966J} {\bibfield  {journal} {\bibinfo
  {journal} {Journal of Materials Chemistry C}\ }\textbf {\bibinfo {volume}
  {9}},\ \bibinfo {pages} {8051} (\bibinfo {year} {2021})}\BibitemShut
  {NoStop}%
\bibitem [{\citenamefont {Ovsyannikov}\ \emph {et~al.}(2021)\citenamefont
  {Ovsyannikov}, \citenamefont {Aslandukova}, \citenamefont {Aslandukov},
  \citenamefont {Chariton}, \citenamefont {Tsirlin}, \citenamefont
  {Korobeynikov}, \citenamefont {Morozova}, \citenamefont {Fedotenko},
  \citenamefont {Khandarkhaeva},\ and\ \citenamefont
  {Dubrovinsky}}]{ovsyannikov2021structural}%
  \BibitemOpen
  \bibfield  {author} {\bibinfo {author} {\bibfnamefont {S.~V.}\ \bibnamefont
  {Ovsyannikov}}, \bibinfo {author} {\bibfnamefont {A.~A.}\ \bibnamefont
  {Aslandukova}}, \bibinfo {author} {\bibfnamefont {A.}~\bibnamefont
  {Aslandukov}}, \bibinfo {author} {\bibfnamefont {S.}~\bibnamefont
  {Chariton}}, \bibinfo {author} {\bibfnamefont {A.~A.}\ \bibnamefont
  {Tsirlin}}, \bibinfo {author} {\bibfnamefont {I.~V.}\ \bibnamefont
  {Korobeynikov}}, \bibinfo {author} {\bibfnamefont {N.~V.}\ \bibnamefont
  {Morozova}}, \bibinfo {author} {\bibfnamefont {T.}~\bibnamefont {Fedotenko}},
  \bibinfo {author} {\bibfnamefont {S.}~\bibnamefont {Khandarkhaeva}},\ and\
  \bibinfo {author} {\bibfnamefont {L.}~\bibnamefont {Dubrovinsky}},\
  }\bibfield  {title} {\bibinfo {title} {Structural stability and properties of
  marokite-type $\gamma$-{M}n$_3${O}$_4$},\ }\href
  {https://pubs.acs.org/doi/full/10.1021/acs.inorgchem.1c01782} {\bibfield
  {journal} {\bibinfo  {journal} {Inorganic Chemistry}\ }\textbf {\bibinfo
  {volume} {60}},\ \bibinfo {pages} {13440} (\bibinfo {year}
  {2021})}\BibitemShut {NoStop}%
\bibitem [{\citenamefont {Nagle-Cocco}\ \emph {et~al.}(2022)\citenamefont
  {Nagle-Cocco}, \citenamefont {Bull}, \citenamefont {Ridley},\ and\
  \citenamefont {Dutton}}]{nagle2022pressure}%
  \BibitemOpen
  \bibfield  {author} {\bibinfo {author} {\bibfnamefont {L.~A.~V.}\
  \bibnamefont {Nagle-Cocco}}, \bibinfo {author} {\bibfnamefont {C.~L.}\
  \bibnamefont {Bull}}, \bibinfo {author} {\bibfnamefont {C.~J.}\ \bibnamefont
  {Ridley}},\ and\ \bibinfo {author} {\bibfnamefont {S.~E.}\ \bibnamefont
  {Dutton}},\ }\bibfield  {title} {\bibinfo {title} {Pressure tuning the
  {J}ahn--{T}eller transition temperature in {N}a{N}i{O}$_2$},\ }\href
  {https://pubs.acs.org/doi/full/10.1021/acs.inorgchem.1c03345} {\bibfield
  {journal} {\bibinfo  {journal} {ACS Inorganic Chemistry}\ }\textbf {\bibinfo
  {volume} {61}},\ \bibinfo {pages} {4312} (\bibinfo {year}
  {2022})}\BibitemShut {NoStop}%
\bibitem [{\citenamefont {Bostr{\"o}m}\ \emph {et~al.}(2024)\citenamefont
  {Bostr{\"o}m}, \citenamefont {Cairns}, \citenamefont {Chen}, \citenamefont
  {Daisenberger}, \citenamefont {Ridley},\ and\ \citenamefont
  {Funnell}}]{bostrom2024pressure}%
  \BibitemOpen
  \bibfield  {author} {\bibinfo {author} {\bibfnamefont {H.~L.~B.}\
  \bibnamefont {Bostr{\"o}m}}, \bibinfo {author} {\bibfnamefont {A.~B.}\
  \bibnamefont {Cairns}}, \bibinfo {author} {\bibfnamefont {M.}~\bibnamefont
  {Chen}}, \bibinfo {author} {\bibfnamefont {D.}~\bibnamefont {Daisenberger}},
  \bibinfo {author} {\bibfnamefont {C.~J.}\ \bibnamefont {Ridley}},\ and\
  \bibinfo {author} {\bibfnamefont {N.~P.}\ \bibnamefont {Funnell}},\
  }\bibfield  {title} {\bibinfo {title} {The pressure response of
  {Jahn--Teller}-distorted {Prussian} blue analogues},\ }\href
  {https://pubs.rsc.org/en/content/articlehtml/2024/sc/d3sc06912e} {\bibfield
  {journal} {\bibinfo  {journal} {Chemical Science}\ }\textbf {\bibinfo
  {volume} {15}},\ \bibinfo {pages} {3155} (\bibinfo {year}
  {2024})}\BibitemShut {NoStop}%
\bibitem [{\citenamefont {Choi}\ \emph {et~al.}(2006)\citenamefont {Choi},
  \citenamefont {Shim},\ and\ \citenamefont {Min}}]{choi2006electronic}%
  \BibitemOpen
  \bibfield  {author} {\bibinfo {author} {\bibfnamefont {H.~C.}\ \bibnamefont
  {Choi}}, \bibinfo {author} {\bibfnamefont {J.~H.}\ \bibnamefont {Shim}},\
  and\ \bibinfo {author} {\bibfnamefont {B.~I.}\ \bibnamefont {Min}},\
  }\bibfield  {title} {\bibinfo {title} {Electronic structures and magnetic
  properties of spinel {Z}n{M}n$_2${O}$_4$ under high pressure},\ }\href
  {https://journals.aps.org/prb/abstract/10.1103/PhysRevB.74.172103} {\bibfield
   {journal} {\bibinfo  {journal} {Phys. Rev. B}\ }\textbf {\bibinfo {volume}
  {74}},\ \bibinfo {pages} {172103} (\bibinfo {year} {2006})}\BibitemShut
  {NoStop}%
\bibitem [{\citenamefont {Dyer}\ \emph {et~al.}(1954)\citenamefont {Dyer},
  \citenamefont {Borie~Jr},\ and\ \citenamefont {Smith}}]{dyer1954alkali}%
  \BibitemOpen
  \bibfield  {author} {\bibinfo {author} {\bibfnamefont {L.~D.}\ \bibnamefont
  {Dyer}}, \bibinfo {author} {\bibfnamefont {B.~S.}\ \bibnamefont {Borie~Jr}},\
  and\ \bibinfo {author} {\bibfnamefont {G.~P.}\ \bibnamefont {Smith}},\
  }\bibfield  {title} {\bibinfo {title} {Alkali metal-nickel oxides of the type
  {MNiO}$_2$},\ }\href {https://pubs.acs.org/doi/pdf/10.1021/ja01635a012}
  {\bibfield  {journal} {\bibinfo  {journal} {Journal of the American Chemical
  Society}\ }\textbf {\bibinfo {volume} {76}},\ \bibinfo {pages} {1499}
  (\bibinfo {year} {1954})}\BibitemShut {NoStop}%
\bibitem [{\citenamefont {Dick}\ \emph {et~al.}(1997)\citenamefont {Dick},
  \citenamefont {M{\"u}ller}, \citenamefont {Preissinger},\ and\ \citenamefont
  {Zeiske}}]{dick1997structure}%
  \BibitemOpen
  \bibfield  {author} {\bibinfo {author} {\bibfnamefont {S.}~\bibnamefont
  {Dick}}, \bibinfo {author} {\bibfnamefont {M.}~\bibnamefont {M{\"u}ller}},
  \bibinfo {author} {\bibfnamefont {F.}~\bibnamefont {Preissinger}},\ and\
  \bibinfo {author} {\bibfnamefont {T.}~\bibnamefont {Zeiske}},\ }\bibfield
  {title} {\bibinfo {title} {The structure of monoclinic {NaNiO}$_2$ as
  determined by powder x-ray and neutron scattering},\ }\href
  {https://www.cambridge.org/core/journals/powder-diffraction/article/structure-of-monoclinic-nanio2-as-determined-by-powder-xray-and-neutron-scattering/497D3D5ABA30A08BE3B41FB89BA9DF0B}
  {\bibfield  {journal} {\bibinfo  {journal} {Powder Diffraction}\ }\textbf
  {\bibinfo {volume} {12}},\ \bibinfo {pages} {239} (\bibinfo {year}
  {1997})}\BibitemShut {NoStop}%
\bibitem [{\citenamefont {Chappel}\ \emph {et~al.}(2000)\citenamefont
  {Chappel}, \citenamefont {Nunez-Regueiro}, \citenamefont {Chouteau},
  \citenamefont {Isnard},\ and\ \citenamefont {Darie}}]{chappel2000study}%
  \BibitemOpen
  \bibfield  {author} {\bibinfo {author} {\bibfnamefont {E.}~\bibnamefont
  {Chappel}}, \bibinfo {author} {\bibfnamefont {M.}~\bibnamefont
  {Nunez-Regueiro}}, \bibinfo {author} {\bibfnamefont {G.}~\bibnamefont
  {Chouteau}}, \bibinfo {author} {\bibfnamefont {O.}~\bibnamefont {Isnard}},\
  and\ \bibinfo {author} {\bibfnamefont {C.}~\bibnamefont {Darie}},\ }\bibfield
   {title} {\bibinfo {title} {Study of the ferrodistorsive orbital ordering in
  {NaNiO}$_2$ by neutron diffraction and submillimeter wave {ESR}},\ }\href
  {https://link.springer.com/article/10.1007/s100510070099} {\bibfield
  {journal} {\bibinfo  {journal} {The European Physical Journal B-Condensed
  Matter and Complex Systems}\ }\textbf {\bibinfo {volume} {17}},\ \bibinfo
  {pages} {615} (\bibinfo {year} {2000})}\BibitemShut {NoStop}%
\bibitem [{\citenamefont {Sofin}\ and\ \citenamefont
  {Jansen}(2005)}]{sofin2005new}%
  \BibitemOpen
  \bibfield  {author} {\bibinfo {author} {\bibfnamefont {M.}~\bibnamefont
  {Sofin}}\ and\ \bibinfo {author} {\bibfnamefont {M.}~\bibnamefont {Jansen}},\
  }\bibfield  {title} {\bibinfo {title} {New route of preparation and
  properties of {NaNiO}$_2$},\ }\href
  {https://www.degruyter.com/document/doi/10.1515/znb-2005-0615/html}
  {\bibfield  {journal} {\bibinfo  {journal} {Zeitschrift f{\"u}r
  Naturforschung B}\ }\textbf {\bibinfo {volume} {60}},\ \bibinfo {pages} {701}
  (\bibinfo {year} {2005})}\BibitemShut {NoStop}%
\bibitem [{\citenamefont {Nagle-Cocco}\ \emph {et~al.}(2024)\citenamefont
  {Nagle-Cocco}, \citenamefont {Genreith-Schriever}, \citenamefont {Steele},
  \citenamefont {Tacconis}, \citenamefont {Bocarsly}, \citenamefont {Mathon},
  \citenamefont {Neuefeind}, \citenamefont {Liu}, \citenamefont {O'Keefe},
  \citenamefont {Goodwin}, \citenamefont {Grey}, \citenamefont {Evans},\ and\
  \citenamefont {Dutton}}]{nagle2024displacive}%
  \BibitemOpen
  \bibfield  {author} {\bibinfo {author} {\bibfnamefont {L.~A.~V.}\
  \bibnamefont {Nagle-Cocco}}, \bibinfo {author} {\bibfnamefont {A.~R.}\
  \bibnamefont {Genreith-Schriever}}, \bibinfo {author} {\bibfnamefont
  {J.~M.~A.}\ \bibnamefont {Steele}}, \bibinfo {author} {\bibfnamefont
  {C.}~\bibnamefont {Tacconis}}, \bibinfo {author} {\bibfnamefont {J.~D.}\
  \bibnamefont {Bocarsly}}, \bibinfo {author} {\bibfnamefont {O.}~\bibnamefont
  {Mathon}}, \bibinfo {author} {\bibfnamefont {J.~C.}\ \bibnamefont
  {Neuefeind}}, \bibinfo {author} {\bibfnamefont {J.}~\bibnamefont {Liu}},
  \bibinfo {author} {\bibfnamefont {C.~A.}\ \bibnamefont {O'Keefe}}, \bibinfo
  {author} {\bibfnamefont {A.~L.}\ \bibnamefont {Goodwin}}, \bibinfo {author}
  {\bibfnamefont {C.~P.}\ \bibnamefont {Grey}}, \bibinfo {author}
  {\bibfnamefont {J.~S.~O.}\ \bibnamefont {Evans}},\ and\ \bibinfo {author}
  {\bibfnamefont {S.~E.}\ \bibnamefont {Dutton}},\ }\bibfield  {title}
  {\bibinfo {title} {Displacive {Jahn--Teller} transition in {NaNiO}$_2$},\
  }\href {https://pubs.acs.org/doi/full/10.1021/jacs.4c09922} {\bibfield
  {journal} {\bibinfo  {journal} {Journal of the American Chemical Society}\
  }\textbf {\bibinfo {volume} {146}},\ \bibinfo {pages} {29560} (\bibinfo
  {year} {2024})}\BibitemShut {NoStop}%
\bibitem [{\citenamefont {Radin}\ \emph {et~al.}(2020)\citenamefont {Radin},
  \citenamefont {Thomas},\ and\ \citenamefont {Van~der Ven}}]{radin2020order}%
  \BibitemOpen
  \bibfield  {author} {\bibinfo {author} {\bibfnamefont {M.~D.}\ \bibnamefont
  {Radin}}, \bibinfo {author} {\bibfnamefont {J.~C.}\ \bibnamefont {Thomas}},\
  and\ \bibinfo {author} {\bibfnamefont {A.}~\bibnamefont {Van~der Ven}},\
  }\bibfield  {title} {\bibinfo {title} {Order-disorder versus displacive
  transitions in {J}ahn--{T}eller active layered materials},\ }\href
  {https://journals.aps.org/prmaterials/abstract/10.1103/PhysRevMaterials.4.043601}
  {\bibfield  {journal} {\bibinfo  {journal} {Physical Review Materials}\
  }\textbf {\bibinfo {volume} {4}},\ \bibinfo {pages} {043601} (\bibinfo {year}
  {2020})}\BibitemShut {NoStop}%
\bibitem [{\citenamefont {Genreith-Schriever}\ \emph
  {et~al.}(2024)\citenamefont {Genreith-Schriever}, \citenamefont {Alexiu},
  \citenamefont {Phillips}, \citenamefont {Coates}, \citenamefont
  {Nagle-Cocco}, \citenamefont {Bocarsly}, \citenamefont {Sayed}, \citenamefont
  {Dutton},\ and\ \citenamefont {Grey}}]{genreith2024jahn}%
  \BibitemOpen
  \bibfield  {author} {\bibinfo {author} {\bibfnamefont {A.~R.}\ \bibnamefont
  {Genreith-Schriever}}, \bibinfo {author} {\bibfnamefont {A.}~\bibnamefont
  {Alexiu}}, \bibinfo {author} {\bibfnamefont {G.~S.}\ \bibnamefont
  {Phillips}}, \bibinfo {author} {\bibfnamefont {C.~S.}\ \bibnamefont
  {Coates}}, \bibinfo {author} {\bibfnamefont {L.~A.~V.}\ \bibnamefont
  {Nagle-Cocco}}, \bibinfo {author} {\bibfnamefont {J.~D.}\ \bibnamefont
  {Bocarsly}}, \bibinfo {author} {\bibfnamefont {F.~N.}\ \bibnamefont {Sayed}},
  \bibinfo {author} {\bibfnamefont {S.~E.}\ \bibnamefont {Dutton}},\ and\
  \bibinfo {author} {\bibfnamefont {C.~P.}\ \bibnamefont {Grey}},\ }\bibfield
  {title} {\bibinfo {title} {Jahn--{T}eller distortions and phase transitions
  in {LiNiO}$_2$: Insights from \textit{ab initio} molecular dynamics and
  variable-temperature x-ray diffraction},\ }\href
  {https://pubs.acs.org/doi/full/10.1021/acs.chemmater.3c02413} {\bibfield
  {journal} {\bibinfo  {journal} {Chemistry of Materials}\ }\textbf {\bibinfo
  {volume} {36}},\ \bibinfo {pages} {2289} (\bibinfo {year}
  {2024})}\BibitemShut {NoStop}%
\bibitem [{\citenamefont {Vassilaras}\ \emph {et~al.}(2012)\citenamefont
  {Vassilaras}, \citenamefont {Ma}, \citenamefont {Li},\ and\ \citenamefont
  {Ceder}}]{vassilaras2012electrochemical}%
  \BibitemOpen
  \bibfield  {author} {\bibinfo {author} {\bibfnamefont {P.}~\bibnamefont
  {Vassilaras}}, \bibinfo {author} {\bibfnamefont {X.}~\bibnamefont {Ma}},
  \bibinfo {author} {\bibfnamefont {X.}~\bibnamefont {Li}},\ and\ \bibinfo
  {author} {\bibfnamefont {G.}~\bibnamefont {Ceder}},\ }\bibfield  {title}
  {\bibinfo {title} {Electrochemical properties of monoclinic {NaNiO}$_2$},\
  }\href {https://iopscience.iop.org/article/10.1149/2.023302jes/meta}
  {\bibfield  {journal} {\bibinfo  {journal} {Journal of The Electrochemical
  Society}\ }\textbf {\bibinfo {volume} {160}},\ \bibinfo {pages} {A207}
  (\bibinfo {year} {2012})}\BibitemShut {NoStop}%
\bibitem [{\citenamefont {Han}\ \emph {et~al.}(2014)\citenamefont {Han},
  \citenamefont {Gonzalo}, \citenamefont {Casas-Cabanas},\ and\ \citenamefont
  {Rojo}}]{han2014structural}%
  \BibitemOpen
  \bibfield  {author} {\bibinfo {author} {\bibfnamefont {M.~H.}\ \bibnamefont
  {Han}}, \bibinfo {author} {\bibfnamefont {E.}~\bibnamefont {Gonzalo}},
  \bibinfo {author} {\bibfnamefont {M.}~\bibnamefont {Casas-Cabanas}},\ and\
  \bibinfo {author} {\bibfnamefont {T.}~\bibnamefont {Rojo}},\ }\bibfield
  {title} {\bibinfo {title} {Structural evolution and electrochemistry of
  monoclinic {NaNiO}$_2$ upon the first cycling process},\ }\href
  {https://www.sciencedirect.com/science/article/pii/S0378775314002328}
  {\bibfield  {journal} {\bibinfo  {journal} {Journal of Power Sources}\
  }\textbf {\bibinfo {volume} {258}},\ \bibinfo {pages} {266} (\bibinfo {year}
  {2014})}\BibitemShut {NoStop}%
\bibitem [{\citenamefont {Wang}\ \emph {et~al.}(2017)\citenamefont {Wang},
  \citenamefont {Wang}, \citenamefont {Zhang}, \citenamefont {Ren},
  \citenamefont {Zuo}, \citenamefont {Yin},\ and\ \citenamefont
  {Wang}}]{wang2017unravelling}%
  \BibitemOpen
  \bibfield  {author} {\bibinfo {author} {\bibfnamefont {L.}~\bibnamefont
  {Wang}}, \bibinfo {author} {\bibfnamefont {J.}~\bibnamefont {Wang}}, \bibinfo
  {author} {\bibfnamefont {X.}~\bibnamefont {Zhang}}, \bibinfo {author}
  {\bibfnamefont {Y.}~\bibnamefont {Ren}}, \bibinfo {author} {\bibfnamefont
  {P.}~\bibnamefont {Zuo}}, \bibinfo {author} {\bibfnamefont {G.}~\bibnamefont
  {Yin}},\ and\ \bibinfo {author} {\bibfnamefont {J.}~\bibnamefont {Wang}},\
  }\bibfield  {title} {\bibinfo {title} {Unravelling the origin of irreversible
  capacity loss in {NaNiO}$_2$ for high voltage sodium ion batteries},\ }\href
  {https://www.sciencedirect.com/science/article/pii/S2211285517301222}
  {\bibfield  {journal} {\bibinfo  {journal} {Nano Energy}\ }\textbf {\bibinfo
  {volume} {34}},\ \bibinfo {pages} {215} (\bibinfo {year} {2017})}\BibitemShut
  {NoStop}%
\bibitem [{\citenamefont {Sada}\ \emph {et~al.}(2024)\citenamefont {Sada},
  \citenamefont {Kmiec},\ and\ \citenamefont {Manthiram}}]{sada2024mitigating}%
  \BibitemOpen
  \bibfield  {author} {\bibinfo {author} {\bibfnamefont {K.}~\bibnamefont
  {Sada}}, \bibinfo {author} {\bibfnamefont {S.}~\bibnamefont {Kmiec}},\ and\
  \bibinfo {author} {\bibfnamefont {A.}~\bibnamefont {Manthiram}},\ }\bibfield
  {title} {\bibinfo {title} {Mitigating sodium ordering for enhanced solid
  solution behavior in layered {NaNiO}$_2$ cathodes},\ }\href
  {https://onlinelibrary.wiley.com/doi/full/10.1002/ange.202403865} {\bibfield
  {journal} {\bibinfo  {journal} {Angewandte Chemie}\ }\textbf {\bibinfo
  {volume} {63}},\ \bibinfo {pages} {e202403865} (\bibinfo {year}
  {2024})}\BibitemShut {NoStop}%
\bibitem [{\citenamefont {Steele}\ \emph {et~al.}(2025)\citenamefont {Steele},
  \citenamefont {Genreith-Schriever}, \citenamefont {Bocarsly}, \citenamefont
  {Nagle-Cocco}, \citenamefont {Sayed}, \citenamefont {Juramy}, \citenamefont
  {O'Keefe}, \citenamefont {Orlandi}, \citenamefont {Manuel}, \citenamefont
  {Dutton},\ and\ \citenamefont {Grey}}]{steele2025structural}%
  \BibitemOpen
  \bibfield  {author} {\bibinfo {author} {\bibfnamefont {J.~M.~A.}\
  \bibnamefont {Steele}}, \bibinfo {author} {\bibfnamefont {A.~R.}\
  \bibnamefont {Genreith-Schriever}}, \bibinfo {author} {\bibfnamefont {J.~D.}\
  \bibnamefont {Bocarsly}}, \bibinfo {author} {\bibfnamefont {L.~A.~V.}\
  \bibnamefont {Nagle-Cocco}}, \bibinfo {author} {\bibfnamefont {F.~N.}\
  \bibnamefont {Sayed}}, \bibinfo {author} {\bibfnamefont {M.}~\bibnamefont
  {Juramy}}, \bibinfo {author} {\bibfnamefont {C.~A.}\ \bibnamefont {O'Keefe}},
  \bibinfo {author} {\bibfnamefont {F.}~\bibnamefont {Orlandi}}, \bibinfo
  {author} {\bibfnamefont {P.}~\bibnamefont {Manuel}}, \bibinfo {author}
  {\bibfnamefont {S.~E.}\ \bibnamefont {Dutton}},\ and\ \bibinfo {author}
  {\bibfnamefont {C.~P.}\ \bibnamefont {Grey}},\ }\bibfield  {title} {\bibinfo
  {title} {Structural elucidation of {N}a$_{2/3}${NiO}$_2$, a dynamically
  stabilised cathode phase with nickel charge- and sodium vacancy ordering},\
  }\href {https://pubs.acs.org/doi/full/10.1021/acs.chemmater.5c00084}
  {\bibfield  {journal} {\bibinfo  {journal} {Chemistry of Materials}\ }
  (\bibinfo {year} {2025})}\BibitemShut {NoStop}%
\bibitem [{\citenamefont {Booth}\ \emph {et~al.}(2021)\citenamefont {Booth},
  \citenamefont {Nedoma}, \citenamefont {Anthonisamy}, \citenamefont {Baker},
  \citenamefont {Boston}, \citenamefont {Bronstein}, \citenamefont {Clarke},
  \citenamefont {Cussen}, \citenamefont {Daramalla}, \citenamefont {De~Volder},
  \citenamefont {Dutton}, \citenamefont {Falkowski}, \citenamefont {Fleck},
  \citenamefont {Geddes}, \citenamefont {Gollapally}, \citenamefont {Goodwin},
  \citenamefont {Griffin}, \citenamefont {Haworth}, \citenamefont {Hayward},
  \citenamefont {Hull}, \citenamefont {Inkson}, \citenamefont {Johnston},
  \citenamefont {Lu}, \citenamefont {MacManus-Driscoll}, \citenamefont
  {Labalde}, \citenamefont {McClelland}, \citenamefont {McCombie},
  \citenamefont {Murdock}, \citenamefont {Nayak}, \citenamefont {Park},
  \citenamefont {Pérez}, \citenamefont {Pickard}, \citenamefont {Piper},
  \citenamefont {Playford}, \citenamefont {Price}, \citenamefont {Scanlon},
  \citenamefont {Stallard}, \citenamefont {Tapia-Ruiz}, \citenamefont {West},
  \citenamefont {Wheatcroft}, \citenamefont {Wilson}, \citenamefont {Zhang},
  \citenamefont {Zhi}, \citenamefont {Zhu},\ and\ \citenamefont
  {Cussen}}]{booth2021perspectives}%
  \BibitemOpen
  \bibfield  {author} {\bibinfo {author} {\bibfnamefont {S.~G.}\ \bibnamefont
  {Booth}}, \bibinfo {author} {\bibfnamefont {A.~J.}\ \bibnamefont {Nedoma}},
  \bibinfo {author} {\bibfnamefont {N.~N.}\ \bibnamefont {Anthonisamy}},
  \bibinfo {author} {\bibfnamefont {P.~J.}\ \bibnamefont {Baker}}, \bibinfo
  {author} {\bibfnamefont {R.}~\bibnamefont {Boston}}, \bibinfo {author}
  {\bibfnamefont {H.}~\bibnamefont {Bronstein}}, \bibinfo {author}
  {\bibfnamefont {S.~J.}\ \bibnamefont {Clarke}}, \bibinfo {author}
  {\bibfnamefont {E.~J.}\ \bibnamefont {Cussen}}, \bibinfo {author}
  {\bibfnamefont {V.}~\bibnamefont {Daramalla}}, \bibinfo {author}
  {\bibfnamefont {M.}~\bibnamefont {De~Volder}}, \bibinfo {author}
  {\bibfnamefont {S.~E.}\ \bibnamefont {Dutton}}, \bibinfo {author}
  {\bibfnamefont {V.}~\bibnamefont {Falkowski}}, \bibinfo {author}
  {\bibfnamefont {N.~A.}\ \bibnamefont {Fleck}}, \bibinfo {author}
  {\bibfnamefont {H.~S.}\ \bibnamefont {Geddes}}, \bibinfo {author}
  {\bibfnamefont {N.}~\bibnamefont {Gollapally}}, \bibinfo {author}
  {\bibfnamefont {A.~L.}\ \bibnamefont {Goodwin}}, \bibinfo {author}
  {\bibfnamefont {J.~M.}\ \bibnamefont {Griffin}}, \bibinfo {author}
  {\bibfnamefont {A.~R.}\ \bibnamefont {Haworth}}, \bibinfo {author}
  {\bibfnamefont {M.~A.}\ \bibnamefont {Hayward}}, \bibinfo {author}
  {\bibfnamefont {S.}~\bibnamefont {Hull}}, \bibinfo {author} {\bibfnamefont
  {B.~J.}\ \bibnamefont {Inkson}}, \bibinfo {author} {\bibfnamefont {B.~J.}\
  \bibnamefont {Johnston}}, \bibinfo {author} {\bibfnamefont {Z.}~\bibnamefont
  {Lu}}, \bibinfo {author} {\bibfnamefont {J.~L.}\ \bibnamefont
  {MacManus-Driscoll}}, \bibinfo {author} {\bibfnamefont {X.~M. D.~I.}\
  \bibnamefont {Labalde}}, \bibinfo {author} {\bibfnamefont {I.}~\bibnamefont
  {McClelland}}, \bibinfo {author} {\bibfnamefont {K.}~\bibnamefont
  {McCombie}}, \bibinfo {author} {\bibfnamefont {B.}~\bibnamefont {Murdock}},
  \bibinfo {author} {\bibfnamefont {D.}~\bibnamefont {Nayak}}, \bibinfo
  {author} {\bibfnamefont {S.}~\bibnamefont {Park}}, \bibinfo {author}
  {\bibfnamefont {G.~E.}\ \bibnamefont {Pérez}}, \bibinfo {author}
  {\bibfnamefont {C.~J.}\ \bibnamefont {Pickard}}, \bibinfo {author}
  {\bibfnamefont {L.~F.~J.}\ \bibnamefont {Piper}}, \bibinfo {author}
  {\bibfnamefont {H.~Y.}\ \bibnamefont {Playford}}, \bibinfo {author}
  {\bibfnamefont {S.}~\bibnamefont {Price}}, \bibinfo {author} {\bibfnamefont
  {D.~O.}\ \bibnamefont {Scanlon}}, \bibinfo {author} {\bibfnamefont {J.~C.}\
  \bibnamefont {Stallard}}, \bibinfo {author} {\bibfnamefont {N.}~\bibnamefont
  {Tapia-Ruiz}}, \bibinfo {author} {\bibfnamefont {A.~R.}\ \bibnamefont
  {West}}, \bibinfo {author} {\bibfnamefont {L.}~\bibnamefont {Wheatcroft}},
  \bibinfo {author} {\bibfnamefont {M.}~\bibnamefont {Wilson}}, \bibinfo
  {author} {\bibfnamefont {L.}~\bibnamefont {Zhang}}, \bibinfo {author}
  {\bibfnamefont {X.}~\bibnamefont {Zhi}}, \bibinfo {author} {\bibfnamefont
  {B.}~\bibnamefont {Zhu}},\ and\ \bibinfo {author} {\bibfnamefont {S.~A.}\
  \bibnamefont {Cussen}},\ }\bibfield  {title} {\bibinfo {title} {Perspectives
  for next generation lithium-ion battery cathode materials},\ }\href
  {https://pubs.aip.org/aip/apm/article/9/10/109201/123081} {\bibfield
  {journal} {\bibinfo  {journal} {APL Materials}\ }\textbf {\bibinfo {volume}
  {9}} (\bibinfo {year} {2021})}\BibitemShut {NoStop}%
\bibitem [{\citenamefont {Knebel}\ \emph {et~al.}(2006)\citenamefont {Knebel},
  \citenamefont {Aoki}, \citenamefont {Braithwaite}, \citenamefont {Salce},\
  and\ \citenamefont {Flouquet}}]{knebel2006coexistence}%
  \BibitemOpen
  \bibfield  {author} {\bibinfo {author} {\bibfnamefont {G.}~\bibnamefont
  {Knebel}}, \bibinfo {author} {\bibfnamefont {D.}~\bibnamefont {Aoki}},
  \bibinfo {author} {\bibfnamefont {D.}~\bibnamefont {Braithwaite}}, \bibinfo
  {author} {\bibfnamefont {B.}~\bibnamefont {Salce}},\ and\ \bibinfo {author}
  {\bibfnamefont {J.}~\bibnamefont {Flouquet}},\ }\bibfield  {title} {\bibinfo
  {title} {Coexistence of antiferromagnetism and superconductivity in
  {CeRhIn}$_5$ under high pressure and magnetic field},\ }\href
  {https://journals.aps.org/prb/abstract/10.1103/PhysRevB.74.020501} {\bibfield
   {journal} {\bibinfo  {journal} {Physical Review B—Condensed Matter and
  Materials Physics}\ }\textbf {\bibinfo {volume} {74}},\ \bibinfo {pages}
  {020501} (\bibinfo {year} {2006})}\BibitemShut {NoStop}%
\bibitem [{\citenamefont {Torikachvili}\ \emph {et~al.}(2008)\citenamefont
  {Torikachvili}, \citenamefont {Bud’ko}, \citenamefont {Ni},\ and\
  \citenamefont {Canfield}}]{torikachvili2008pressure}%
  \BibitemOpen
  \bibfield  {author} {\bibinfo {author} {\bibfnamefont {M.~S.}\ \bibnamefont
  {Torikachvili}}, \bibinfo {author} {\bibfnamefont {S.~L.}\ \bibnamefont
  {Bud’ko}}, \bibinfo {author} {\bibfnamefont {N.}~\bibnamefont {Ni}},\ and\
  \bibinfo {author} {\bibfnamefont {P.~C.}\ \bibnamefont {Canfield}},\
  }\bibfield  {title} {\bibinfo {title} {Pressure induced superconductivity in
  {CaFe}$_2${As}$_2$},\ }\href
  {https://journals.aps.org/prl/abstract/10.1103/PhysRevLett.101.057006}
  {\bibfield  {journal} {\bibinfo  {journal} {Physical Review Letters}\
  }\textbf {\bibinfo {volume} {101}},\ \bibinfo {pages} {057006} (\bibinfo
  {year} {2008})}\BibitemShut {NoStop}%
\bibitem [{\citenamefont {Pan}\ \emph {et~al.}(2015)\citenamefont {Pan},
  \citenamefont {Chen}, \citenamefont {Liu}, \citenamefont {Feng},
  \citenamefont {Wei}, \citenamefont {Zhou}, \citenamefont {Chi}, \citenamefont
  {Pi}, \citenamefont {Yen}, \citenamefont {Song} \emph
  {et~al.}}]{pan2015pressure}%
  \BibitemOpen
  \bibfield  {author} {\bibinfo {author} {\bibfnamefont {X.-C.}\ \bibnamefont
  {Pan}}, \bibinfo {author} {\bibfnamefont {X.}~\bibnamefont {Chen}}, \bibinfo
  {author} {\bibfnamefont {H.}~\bibnamefont {Liu}}, \bibinfo {author}
  {\bibfnamefont {Y.}~\bibnamefont {Feng}}, \bibinfo {author} {\bibfnamefont
  {Z.}~\bibnamefont {Wei}}, \bibinfo {author} {\bibfnamefont {Y.}~\bibnamefont
  {Zhou}}, \bibinfo {author} {\bibfnamefont {Z.}~\bibnamefont {Chi}}, \bibinfo
  {author} {\bibfnamefont {L.}~\bibnamefont {Pi}}, \bibinfo {author}
  {\bibfnamefont {F.}~\bibnamefont {Yen}}, \bibinfo {author} {\bibfnamefont
  {F.}~\bibnamefont {Song}}, \emph {et~al.},\ }\bibfield  {title} {\bibinfo
  {title} {Pressure-driven dome-shaped superconductivity and electronic
  structural evolution in tungsten ditelluride},\ }\href
  {https://www.nature.com/articles/ncomms8805} {\bibfield  {journal} {\bibinfo
  {journal} {Nature Communications}\ }\textbf {\bibinfo {volume} {6}},\
  \bibinfo {pages} {7805} (\bibinfo {year} {2015})}\BibitemShut {NoStop}%
\bibitem [{\citenamefont {Wang}\ \emph {et~al.}(2023)\citenamefont {Wang},
  \citenamefont {Wang}, \citenamefont {An}, \citenamefont {Zhou}, \citenamefont
  {Zhou}, \citenamefont {Chen}, \citenamefont {Hao},\ and\ \citenamefont
  {Yang}}]{wang2023two}%
  \BibitemOpen
  \bibfield  {author} {\bibinfo {author} {\bibfnamefont {S.}~\bibnamefont
  {Wang}}, \bibinfo {author} {\bibfnamefont {Q.}~\bibnamefont {Wang}}, \bibinfo
  {author} {\bibfnamefont {C.}~\bibnamefont {An}}, \bibinfo {author}
  {\bibfnamefont {Y.}~\bibnamefont {Zhou}}, \bibinfo {author} {\bibfnamefont
  {Y.}~\bibnamefont {Zhou}}, \bibinfo {author} {\bibfnamefont {X.}~\bibnamefont
  {Chen}}, \bibinfo {author} {\bibfnamefont {N.}~\bibnamefont {Hao}},\ and\
  \bibinfo {author} {\bibfnamefont {Z.}~\bibnamefont {Yang}},\ }\bibfield
  {title} {\bibinfo {title} {Two distinct charge density wave orders and
  emergent superconductivity in pressurized {CuTe}},\ }\href
  {https://www.cell.com/matter/abstract/S2590-2385(23)00376-4} {\bibfield
  {journal} {\bibinfo  {journal} {Matter}\ }\textbf {\bibinfo {volume} {6}},\
  \bibinfo {pages} {3526} (\bibinfo {year} {2023})}\BibitemShut {NoStop}%
\bibitem [{\citenamefont {Anzellini}\ \emph {et~al.}(2018)\citenamefont
  {Anzellini}, \citenamefont {Kleppe}, \citenamefont {Daisenberger},
  \citenamefont {Wharmby}, \citenamefont {Giampaoli}, \citenamefont {Boccato},
  \citenamefont {Baron}, \citenamefont {Miozzi}, \citenamefont {Keeble},
  \citenamefont {Ross}, \citenamefont {Gurney}, \citenamefont {Thompson},
  \citenamefont {Knap}, \citenamefont {Booth}, \citenamefont {Hudson},
  \citenamefont {Hawkins}, \citenamefont {Walter},\ and\ \citenamefont
  {Wilhelm}}]{anzellini2018laser}%
  \BibitemOpen
  \bibfield  {author} {\bibinfo {author} {\bibfnamefont {S.}~\bibnamefont
  {Anzellini}}, \bibinfo {author} {\bibfnamefont {A.~K.}\ \bibnamefont
  {Kleppe}}, \bibinfo {author} {\bibfnamefont {D.}~\bibnamefont
  {Daisenberger}}, \bibinfo {author} {\bibfnamefont {M.~T.}\ \bibnamefont
  {Wharmby}}, \bibinfo {author} {\bibfnamefont {R.}~\bibnamefont {Giampaoli}},
  \bibinfo {author} {\bibfnamefont {S.}~\bibnamefont {Boccato}}, \bibinfo
  {author} {\bibfnamefont {M.~A.}\ \bibnamefont {Baron}}, \bibinfo {author}
  {\bibfnamefont {F.}~\bibnamefont {Miozzi}}, \bibinfo {author} {\bibfnamefont
  {D.~S.}\ \bibnamefont {Keeble}}, \bibinfo {author} {\bibfnamefont
  {A.}~\bibnamefont {Ross}}, \bibinfo {author} {\bibfnamefont {S.}~\bibnamefont
  {Gurney}}, \bibinfo {author} {\bibfnamefont {J.}~\bibnamefont {Thompson}},
  \bibinfo {author} {\bibfnamefont {G.}~\bibnamefont {Knap}}, \bibinfo {author}
  {\bibfnamefont {M.}~\bibnamefont {Booth}}, \bibinfo {author} {\bibfnamefont
  {L.}~\bibnamefont {Hudson}}, \bibinfo {author} {\bibfnamefont
  {D.}~\bibnamefont {Hawkins}}, \bibinfo {author} {\bibfnamefont {M.~J.}\
  \bibnamefont {Walter}},\ and\ \bibinfo {author} {\bibfnamefont
  {H.}~\bibnamefont {Wilhelm}},\ }\bibfield  {title} {\bibinfo {title}
  {Laser-heating system for high-pressure {X}-ray diffraction at the extreme
  conditions beamline {I}15 at {Diamond Light Source}},\ }\href
  {https://journals.iucr.org/s/issues/2018/06/00/il5019/il5019.pdf} {\bibfield
  {journal} {\bibinfo  {journal} {Journal of Synchrotron Radiation}\ }\textbf
  {\bibinfo {volume} {25}},\ \bibinfo {pages} {1860} (\bibinfo {year}
  {2018})}\BibitemShut {NoStop}%
\bibitem [{\citenamefont {Prescher}\ and\ \citenamefont
  {Prakapenka}(2015)}]{prescher2015dioptas}%
  \BibitemOpen
  \bibfield  {author} {\bibinfo {author} {\bibfnamefont {C.}~\bibnamefont
  {Prescher}}\ and\ \bibinfo {author} {\bibfnamefont {V.~B.}\ \bibnamefont
  {Prakapenka}},\ }\bibfield  {title} {\bibinfo {title} {{DIOPTAS}: a program
  for reduction of two-dimensional x-ray diffraction data and data
  exploration},\ }\href
  {https://www.tandfonline.com/doi/full/10.1080/08957959.2015.1059835}
  {\bibfield  {journal} {\bibinfo  {journal} {High Pressure Research}\ }\textbf
  {\bibinfo {volume} {35}},\ \bibinfo {pages} {223} (\bibinfo {year}
  {2015})}\BibitemShut {NoStop}%
\bibitem [{\citenamefont {Coelho}(2018)}]{coelho2018topas}%
  \BibitemOpen
  \bibfield  {author} {\bibinfo {author} {\bibfnamefont {A.~A.}\ \bibnamefont
  {Coelho}},\ }\bibfield  {title} {\bibinfo {title} {{TOPAS} and
  {TOPAS}-{A}cademic: an optimization program integrating computer algebra and
  crystallographic objects written in {C}++},\ }\href
  {https://journals.iucr.org/j/issues/2018/01/00/jo5037/jo5037.pdf} {\bibfield
  {journal} {\bibinfo  {journal} {Journal of Applied Crystallography}\ }\textbf
  {\bibinfo {volume} {51}},\ \bibinfo {pages} {210} (\bibinfo {year}
  {2018})}\BibitemShut {NoStop}%
\bibitem [{\citenamefont {Pawley}(1981)}]{pawley1981unit}%
  \BibitemOpen
  \bibfield  {author} {\bibinfo {author} {\bibfnamefont {G.~S.}\ \bibnamefont
  {Pawley}},\ }\bibfield  {title} {\bibinfo {title} {Unit-cell refinement from
  powder diffraction scans},\ }\href
  {https://journals.iucr.org/paper?pii=S0021889881009618} {\bibfield  {journal}
  {\bibinfo  {journal} {Journal of Applied Crystallography}\ }\textbf {\bibinfo
  {volume} {14}},\ \bibinfo {pages} {357} (\bibinfo {year} {1981})}\BibitemShut
  {NoStop}%
\bibitem [{\citenamefont {Thompson}\ \emph {et~al.}(1987)\citenamefont
  {Thompson}, \citenamefont {Cox},\ and\ \citenamefont
  {Hastings}}]{thompson1987rietveld}%
  \BibitemOpen
  \bibfield  {author} {\bibinfo {author} {\bibfnamefont {P.}~\bibnamefont
  {Thompson}}, \bibinfo {author} {\bibfnamefont {D.~E.}\ \bibnamefont {Cox}},\
  and\ \bibinfo {author} {\bibfnamefont {J.~B.}\ \bibnamefont {Hastings}},\
  }\bibfield  {title} {\bibinfo {title} {Rietveld refinement of
  {Debye--Scherrer} synchrotron x-ray data from {Al$_2$O$_3$}},\ }\href
  {https://journals.iucr.org/j/issues/1987/02/00/a27720/a27720.pdf} {\bibfield
  {journal} {\bibinfo  {journal} {Journal of Applied Crystallography}\ }\textbf
  {\bibinfo {volume} {20}},\ \bibinfo {pages} {79} (\bibinfo {year}
  {1987})}\BibitemShut {NoStop}%
\bibitem [{\citenamefont {Cheary}\ and\ \citenamefont
  {Coelho}(1998)}]{cheary1998axial}%
  \BibitemOpen
  \bibfield  {author} {\bibinfo {author} {\bibfnamefont {R.~W.}\ \bibnamefont
  {Cheary}}\ and\ \bibinfo {author} {\bibfnamefont {A.~A.}\ \bibnamefont
  {Coelho}},\ }\bibfield  {title} {\bibinfo {title} {Axial divergence in a
  conventional {X}-ray powder diffractometer. {I}. {Theoretical} foundations},\
  }\href {https://journals.iucr.org/j/issues/1998/06/00/hw0065/hw0065.pdf}
  {\bibfield  {journal} {\bibinfo  {journal} {Journal of Applied
  Crystallography}\ }\textbf {\bibinfo {volume} {31}},\ \bibinfo {pages} {851}
  (\bibinfo {year} {1998})}\BibitemShut {NoStop}%
\bibitem [{\citenamefont {Rietveld}(1969)}]{rietveld1969profile}%
  \BibitemOpen
  \bibfield  {author} {\bibinfo {author} {\bibfnamefont {H.~M.}\ \bibnamefont
  {Rietveld}},\ }\bibfield  {title} {\bibinfo {title} {A profile refinement
  method for nuclear and magnetic structures},\ }\href
  {https://journals.iucr.org/paper?pii=S0021889869006558} {\bibfield  {journal}
  {\bibinfo  {journal} {Journal of Applied Crystallography}\ }\textbf {\bibinfo
  {volume} {2}},\ \bibinfo {pages} {65} (\bibinfo {year} {1969})}\BibitemShut
  {NoStop}%
\bibitem [{\citenamefont {Birch}(1947)}]{birch1947finite}%
  \BibitemOpen
  \bibfield  {author} {\bibinfo {author} {\bibfnamefont {F.}~\bibnamefont
  {Birch}},\ }\bibfield  {title} {\bibinfo {title} {Finite elastic strain of
  cubic crystals},\ }\href
  {https://journals.aps.org/pr/abstract/10.1103/PhysRev.71.809} {\bibfield
  {journal} {\bibinfo  {journal} {Physical Review}\ }\textbf {\bibinfo {volume}
  {71}},\ \bibinfo {pages} {809} (\bibinfo {year} {1947})}\BibitemShut
  {NoStop}%
\bibitem [{\citenamefont {Angel}(2000)}]{angel2000equations}%
  \BibitemOpen
  \bibfield  {author} {\bibinfo {author} {\bibfnamefont {R.~J.}\ \bibnamefont
  {Angel}},\ }\bibfield  {title} {\bibinfo {title} {Equations of state},\
  }\href
  {https://pubs.geoscienceworld.org/msa/rimg/article/41/1/35/140704/Equations-of-State}
  {\bibfield  {journal} {\bibinfo  {journal} {Reviews in Mineralogy and
  Geochemistry}\ }\textbf {\bibinfo {volume} {41}},\ \bibinfo {pages} {35}
  (\bibinfo {year} {2000})}\BibitemShut {NoStop}%
\bibitem [{\citenamefont {Cliffe}\ and\ \citenamefont
  {Goodwin}(2012)}]{cliffe2012pascal}%
  \BibitemOpen
  \bibfield  {author} {\bibinfo {author} {\bibfnamefont {M.~J.}\ \bibnamefont
  {Cliffe}}\ and\ \bibinfo {author} {\bibfnamefont {A.~L.}\ \bibnamefont
  {Goodwin}},\ }\bibfield  {title} {\bibinfo {title} {P{ASC}al: a principal
  axis strain calculator for thermal expansion and compressibility
  determination},\ }\href
  {https://journals.iucr.org/j/issues/2012/06/00/nb5025/nb5025.pdf} {\bibfield
  {journal} {\bibinfo  {journal} {Journal of Applied Crystallography}\ }\textbf
  {\bibinfo {volume} {45}},\ \bibinfo {pages} {1321} (\bibinfo {year}
  {2012})}\BibitemShut {NoStop}%
\bibitem [{\citenamefont {Lertkiattrakul}\ \emph {et~al.}(2023)\citenamefont
  {Lertkiattrakul}, \citenamefont {Evans},\ and\ \citenamefont
  {Cliffe}}]{lertkiattrakul2023pascal}%
  \BibitemOpen
  \bibfield  {author} {\bibinfo {author} {\bibfnamefont {M.}~\bibnamefont
  {Lertkiattrakul}}, \bibinfo {author} {\bibfnamefont {M.~L.}\ \bibnamefont
  {Evans}},\ and\ \bibinfo {author} {\bibfnamefont {M.~J.}\ \bibnamefont
  {Cliffe}},\ }\bibfield  {title} {\bibinfo {title} {Pascal python: a principal
  axis strain calculator},\ }\href
  {https://joss.theoj.org/papers/10.21105/joss.05556.pdf} {\bibfield  {journal}
  {\bibinfo  {journal} {Journal of Open Source Software}\ }\textbf {\bibinfo
  {volume} {8}},\ \bibinfo {pages} {5556} (\bibinfo {year} {2023})}\BibitemShut
  {NoStop}%
\bibitem [{\citenamefont {Angel}\ and\ \citenamefont
  {Ross}(1996)}]{angel1996compression}%
  \BibitemOpen
  \bibfield  {author} {\bibinfo {author} {\bibfnamefont {R.~J.}\ \bibnamefont
  {Angel}}\ and\ \bibinfo {author} {\bibfnamefont {N.~L.}\ \bibnamefont
  {Ross}},\ }\bibfield  {title} {\bibinfo {title} {Compression mechanisms and
  equations of state},\ }\href
  {https://royalsocietypublishing.org/doi/abs/10.1098/rsta.1996.0057}
  {\bibfield  {journal} {\bibinfo  {journal} {Philosophical Transactions of the
  Royal Society of London. Series A: Mathematical, Physical and Engineering
  Sciences}\ }\textbf {\bibinfo {volume} {354}},\ \bibinfo {pages} {1449}
  (\bibinfo {year} {1996})}\BibitemShut {NoStop}%
\bibitem [{\citenamefont {Katsura}\ and\ \citenamefont
  {Tange}(2019)}]{katsura2019simple}%
  \BibitemOpen
  \bibfield  {author} {\bibinfo {author} {\bibfnamefont {T.}~\bibnamefont
  {Katsura}}\ and\ \bibinfo {author} {\bibfnamefont {Y.}~\bibnamefont
  {Tange}},\ }\bibfield  {title} {\bibinfo {title} {A simple derivation of the
  {Birch--Murnaghan} equations of state ({EOSs}) and comparison with {EOSs}
  derived from other definitions of finite strain},\ }\href
  {https://www.mdpi.com/2075-163X/9/12/745} {\bibfield  {journal} {\bibinfo
  {journal} {Minerals}\ }\textbf {\bibinfo {volume} {9}},\ \bibinfo {pages}
  {745} (\bibinfo {year} {2019})}\BibitemShut {NoStop}%
\bibitem [{\citenamefont {Ballaran}\ and\ \citenamefont
  {Angel}(2003)}]{ballaran2003equation}%
  \BibitemOpen
  \bibfield  {author} {\bibinfo {author} {\bibfnamefont {T.~B.}\ \bibnamefont
  {Ballaran}}\ and\ \bibinfo {author} {\bibfnamefont {R.~J.}\ \bibnamefont
  {Angel}},\ }\bibfield  {title} {\bibinfo {title} {Equation of state and
  high-pressure phase transitions in lawsonite},\ }\href
  {https://pubs.geoscienceworld.org/ejm/eurjmin/article/15/2/241/62009/Equation-of-state-and-high-pressure-phase}
  {\bibfield  {journal} {\bibinfo  {journal} {European Journal of Mineralogy}\
  }\textbf {\bibinfo {volume} {15}},\ \bibinfo {pages} {241} (\bibinfo {year}
  {2003})}\BibitemShut {NoStop}%
\bibitem [{\citenamefont {Pavese}\ \emph {et~al.}(2003)\citenamefont {Pavese},
  \citenamefont {Levy}, \citenamefont {Curetti}, \citenamefont {Diella},
  \citenamefont {Fumagalli},\ and\ \citenamefont {Sani}}]{pavese2003equation}%
  \BibitemOpen
  \bibfield  {author} {\bibinfo {author} {\bibfnamefont {A.}~\bibnamefont
  {Pavese}}, \bibinfo {author} {\bibfnamefont {D.}~\bibnamefont {Levy}},
  \bibinfo {author} {\bibfnamefont {N.}~\bibnamefont {Curetti}}, \bibinfo
  {author} {\bibfnamefont {V.}~\bibnamefont {Diella}}, \bibinfo {author}
  {\bibfnamefont {P.}~\bibnamefont {Fumagalli}},\ and\ \bibinfo {author}
  {\bibfnamefont {A.}~\bibnamefont {Sani}},\ }\bibfield  {title} {\bibinfo
  {title} {Equation of state and compressibility of phlogopite by in-situ
  high-pressure {X}-ray powder diffraction},\ }\href
  {https://pubs.geoscienceworld.org/ejm/eurjmin/article/15/3/455/62032/Equation-of-state-and-compressibility-of}
  {\bibfield  {journal} {\bibinfo  {journal} {European Journal of Mineralogy}\
  }\textbf {\bibinfo {volume} {15}},\ \bibinfo {pages} {455} (\bibinfo {year}
  {2003})}\BibitemShut {NoStop}%
\bibitem [{\citenamefont {Zanazzi}\ \emph {et~al.}(2007)\citenamefont
  {Zanazzi}, \citenamefont {Montagnoli}, \citenamefont {Nazzareni},\ and\
  \citenamefont {Comodi}}]{zanazzi2007structural}%
  \BibitemOpen
  \bibfield  {author} {\bibinfo {author} {\bibfnamefont {P.~F.}\ \bibnamefont
  {Zanazzi}}, \bibinfo {author} {\bibfnamefont {M.}~\bibnamefont {Montagnoli}},
  \bibinfo {author} {\bibfnamefont {S.}~\bibnamefont {Nazzareni}},\ and\
  \bibinfo {author} {\bibfnamefont {P.}~\bibnamefont {Comodi}},\ }\bibfield
  {title} {\bibinfo {title} {Structural effects of pressure on monoclinic
  chlorite: {A} single-crystal study},\ }\href
  {https://pubs.geoscienceworld.org/msa/ammin/article/92/4/655/134449/Structural-effects-of-pressure-on-monoclinic}
  {\bibfield  {journal} {\bibinfo  {journal} {American Mineralogist}\ }\textbf
  {\bibinfo {volume} {92}},\ \bibinfo {pages} {655} (\bibinfo {year}
  {2007})}\BibitemShut {NoStop}%
\bibitem [{\citenamefont {Fujimoto}\ \emph {et~al.}(2018)\citenamefont
  {Fujimoto}, \citenamefont {Akahama}, \citenamefont {Fukui}, \citenamefont
  {Hirao},\ and\ \citenamefont {Ohishi}}]{fujimoto2018observation}%
  \BibitemOpen
  \bibfield  {author} {\bibinfo {author} {\bibfnamefont {M.}~\bibnamefont
  {Fujimoto}}, \bibinfo {author} {\bibfnamefont {Y.}~\bibnamefont {Akahama}},
  \bibinfo {author} {\bibfnamefont {H.}~\bibnamefont {Fukui}}, \bibinfo
  {author} {\bibfnamefont {N.}~\bibnamefont {Hirao}},\ and\ \bibinfo {author}
  {\bibfnamefont {Y.}~\bibnamefont {Ohishi}},\ }\bibfield  {title} {\bibinfo
  {title} {Observation of the negative pressure derivative of the bulk modulus
  in monoclinic {ZrO}$_2$},\ }\href
  {https://pubs.aip.org/aip/adv/article/8/1/015310/991701} {\bibfield
  {journal} {\bibinfo  {journal} {AIP Advances}\ }\textbf {\bibinfo {volume}
  {8}} (\bibinfo {year} {2018})}\BibitemShut {NoStop}%
\bibitem [{\citenamefont {Pavese}(2005)}]{pavese2005relations}%
  \BibitemOpen
  \bibfield  {author} {\bibinfo {author} {\bibfnamefont {A.}~\bibnamefont
  {Pavese}},\ }\bibfield  {title} {\bibinfo {title} {About the relations
  between finite strain in non-cubic crystals and the related phenomenological
  {P-V} {Equation of State}},\ }\href
  {https://link.springer.com/article/10.1007/s00269-005-0465-8} {\bibfield
  {journal} {\bibinfo  {journal} {Physics and Chemistry of Minerals}\ }\textbf
  {\bibinfo {volume} {32}},\ \bibinfo {pages} {269} (\bibinfo {year}
  {2005})}\BibitemShut {NoStop}%
\bibitem [{\citenamefont {Salmani-Rezaie}\ \emph {et~al.}(2020)\citenamefont
  {Salmani-Rezaie}, \citenamefont {Ahadi}, \citenamefont {Strickland},\ and\
  \citenamefont {Stemmer}}]{salmani2020order}%
  \BibitemOpen
  \bibfield  {author} {\bibinfo {author} {\bibfnamefont {S.}~\bibnamefont
  {Salmani-Rezaie}}, \bibinfo {author} {\bibfnamefont {K.}~\bibnamefont
  {Ahadi}}, \bibinfo {author} {\bibfnamefont {W.~M.}\ \bibnamefont
  {Strickland}},\ and\ \bibinfo {author} {\bibfnamefont {S.}~\bibnamefont
  {Stemmer}},\ }\bibfield  {title} {\bibinfo {title} {Order-disorder
  ferroelectric transition of strained {SrTiO}$_3$},\ }\href
  {https://journals.aps.org/prl/abstract/10.1103/PhysRevLett.125.087601}
  {\bibfield  {journal} {\bibinfo  {journal} {Physical Review Letters}\
  }\textbf {\bibinfo {volume} {125}},\ \bibinfo {pages} {087601} (\bibinfo
  {year} {2020})}\BibitemShut {NoStop}%
\bibitem [{\citenamefont {Sicolo}\ \emph {et~al.}(2020)\citenamefont {Sicolo},
  \citenamefont {Mock}, \citenamefont {Bianchini},\ and\ \citenamefont
  {Albe}}]{sicolo2020and}%
  \BibitemOpen
  \bibfield  {author} {\bibinfo {author} {\bibfnamefont {S.}~\bibnamefont
  {Sicolo}}, \bibinfo {author} {\bibfnamefont {M.}~\bibnamefont {Mock}},
  \bibinfo {author} {\bibfnamefont {M.}~\bibnamefont {Bianchini}},\ and\
  \bibinfo {author} {\bibfnamefont {K.}~\bibnamefont {Albe}},\ }\bibfield
  {title} {\bibinfo {title} {And yet it moves: {LiNiO}$_2$, a dynamic
  {J}ahn--{T}eller system},\ }\href
  {https://pubs.acs.org/doi/full/10.1021/acs.chemmater.0c03442} {\bibfield
  {journal} {\bibinfo  {journal} {Chemistry of Materials}\ }\textbf {\bibinfo
  {volume} {32}},\ \bibinfo {pages} {10096} (\bibinfo {year}
  {2020})}\BibitemShut {NoStop}%
\bibitem [{\citenamefont {Rodrigues}\ \emph {et~al.}(2023)\citenamefont
  {Rodrigues}, \citenamefont {Rosa}, \citenamefont {Garbarino}, \citenamefont
  {Irifune}, \citenamefont {Martinez}, \citenamefont {Alonso},\ and\
  \citenamefont {Mathon}}]{rodrigues2023evidence}%
  \BibitemOpen
  \bibfield  {author} {\bibinfo {author} {\bibfnamefont {J.~E.}\ \bibnamefont
  {Rodrigues}}, \bibinfo {author} {\bibfnamefont {A.}~\bibnamefont {Rosa}},
  \bibinfo {author} {\bibfnamefont {G.}~\bibnamefont {Garbarino}}, \bibinfo
  {author} {\bibfnamefont {T.}~\bibnamefont {Irifune}}, \bibinfo {author}
  {\bibfnamefont {J.~L.}\ \bibnamefont {Martinez}}, \bibinfo {author}
  {\bibfnamefont {J.~A.}\ \bibnamefont {Alonso}},\ and\ \bibinfo {author}
  {\bibfnamefont {O.}~\bibnamefont {Mathon}},\ }\bibfield  {title} {\bibinfo
  {title} {Evidence for a pressure-induced phase transition in the highly
  distorted {TlNiO}$_3$ nickelate},\ }\href
  {https://pubs.acs.org/doi/full/10.1021/acs.chemmater.3c00593} {\bibfield
  {journal} {\bibinfo  {journal} {Chemistry of Materials}\ }\textbf {\bibinfo
  {volume} {35}},\ \bibinfo {pages} {5079} (\bibinfo {year}
  {2023})}\BibitemShut {NoStop}%
\bibitem [{\citenamefont {Herlihy}\ \emph {et~al.}(2021)\citenamefont
  {Herlihy}, \citenamefont {Geddes}, \citenamefont {Sosso}, \citenamefont
  {Bull}, \citenamefont {Ridley}, \citenamefont {Goodwin}, \citenamefont
  {Senn},\ and\ \citenamefont {Funnell}}]{herlihy2021recovering}%
  \BibitemOpen
  \bibfield  {author} {\bibinfo {author} {\bibfnamefont {A.}~\bibnamefont
  {Herlihy}}, \bibinfo {author} {\bibfnamefont {H.~S.}\ \bibnamefont {Geddes}},
  \bibinfo {author} {\bibfnamefont {G.~C.}\ \bibnamefont {Sosso}}, \bibinfo
  {author} {\bibfnamefont {C.~L.}\ \bibnamefont {Bull}}, \bibinfo {author}
  {\bibfnamefont {C.~J.}\ \bibnamefont {Ridley}}, \bibinfo {author}
  {\bibfnamefont {A.~L.}\ \bibnamefont {Goodwin}}, \bibinfo {author}
  {\bibfnamefont {M.~S.}\ \bibnamefont {Senn}},\ and\ \bibinfo {author}
  {\bibfnamefont {N.~P.}\ \bibnamefont {Funnell}},\ }\bibfield  {title}
  {\bibinfo {title} {Recovering local structure information from high-pressure
  total scattering experiments},\ }\href
  {https://journals.iucr.org/j/issues/2021/06/00/kc5131/kc5131.pdf} {\bibfield
  {journal} {\bibinfo  {journal} {Journal of Applied Crystallography}\ }\textbf
  {\bibinfo {volume} {54}},\ \bibinfo {pages} {1546} (\bibinfo {year}
  {2021})}\BibitemShut {NoStop}%
\bibitem [{\citenamefont {McHardy}\ \emph {et~al.}(2023)\citenamefont
  {McHardy}, \citenamefont {Storm}, \citenamefont {Duff}, \citenamefont
  {Macleod},\ and\ \citenamefont {McMahon}}]{mchardy2023creation}%
  \BibitemOpen
  \bibfield  {author} {\bibinfo {author} {\bibfnamefont {J.~D.}\ \bibnamefont
  {McHardy}}, \bibinfo {author} {\bibfnamefont {C.~V.}\ \bibnamefont {Storm}},
  \bibinfo {author} {\bibfnamefont {M.~J.}\ \bibnamefont {Duff}}, \bibinfo
  {author} {\bibfnamefont {S.~G.}\ \bibnamefont {Macleod}},\ and\ \bibinfo
  {author} {\bibfnamefont {M.~I.}\ \bibnamefont {McMahon}},\ }\bibfield
  {title} {\bibinfo {title} {On the creation of thermal equations of state for
  use in {Dioptas}},\ }\href
  {https://www.tandfonline.com/doi/full/10.1080/08957959.2023.2187294}
  {\bibfield  {journal} {\bibinfo  {journal} {High Pressure Research}\ }\textbf
  {\bibinfo {volume} {43}},\ \bibinfo {pages} {40} (\bibinfo {year}
  {2023})}\BibitemShut {NoStop}%
\bibitem [{\citenamefont {Shen}\ \emph {et~al.}(2020)\citenamefont {Shen},
  \citenamefont {Wang}, \citenamefont {Dewaele}, \citenamefont {Wu},
  \citenamefont {Fratanduono}, \citenamefont {Eggert}, \citenamefont {Klotz},
  \citenamefont {Dziubek}, \citenamefont {Loubeyre}, \citenamefont
  {Fat’yanov} \emph {et~al.}}]{shen2020toward}%
  \BibitemOpen
  \bibfield  {author} {\bibinfo {author} {\bibfnamefont {G.}~\bibnamefont
  {Shen}}, \bibinfo {author} {\bibfnamefont {Y.}~\bibnamefont {Wang}}, \bibinfo
  {author} {\bibfnamefont {A.}~\bibnamefont {Dewaele}}, \bibinfo {author}
  {\bibfnamefont {C.}~\bibnamefont {Wu}}, \bibinfo {author} {\bibfnamefont
  {D.~E.}\ \bibnamefont {Fratanduono}}, \bibinfo {author} {\bibfnamefont
  {J.}~\bibnamefont {Eggert}}, \bibinfo {author} {\bibfnamefont
  {S.}~\bibnamefont {Klotz}}, \bibinfo {author} {\bibfnamefont {K.~F.}\
  \bibnamefont {Dziubek}}, \bibinfo {author} {\bibfnamefont {P.}~\bibnamefont
  {Loubeyre}}, \bibinfo {author} {\bibfnamefont {O.~V.}\ \bibnamefont
  {Fat’yanov}}, \emph {et~al.},\ }\bibfield  {title} {\bibinfo {title}
  {Toward an international practical pressure scale: A proposal for an {IPPS}
  ruby gauge ({IPPS-Ruby2020})},\ }\href
  {https://www.tandfonline.com/doi/full/10.1080/08957959.2020.1791107}
  {\bibfield  {journal} {\bibinfo  {journal} {High Pressure Research}\ }\textbf
  {\bibinfo {volume} {40}},\ \bibinfo {pages} {299} (\bibinfo {year}
  {2020})}\BibitemShut {NoStop}%
\bibitem [{\citenamefont {Datchi}\ \emph {et~al.}(2007)\citenamefont {Datchi},
  \citenamefont {Dewaele}, \citenamefont {Loubeyre}, \citenamefont {Letoullec},
  \citenamefont {Le~Godec},\ and\ \citenamefont {Canny}}]{datchi2007optical}%
  \BibitemOpen
  \bibfield  {author} {\bibinfo {author} {\bibfnamefont {F.}~\bibnamefont
  {Datchi}}, \bibinfo {author} {\bibfnamefont {A.}~\bibnamefont {Dewaele}},
  \bibinfo {author} {\bibfnamefont {P.}~\bibnamefont {Loubeyre}}, \bibinfo
  {author} {\bibfnamefont {R.}~\bibnamefont {Letoullec}}, \bibinfo {author}
  {\bibfnamefont {Y.}~\bibnamefont {Le~Godec}},\ and\ \bibinfo {author}
  {\bibfnamefont {B.}~\bibnamefont {Canny}},\ }\bibfield  {title} {\bibinfo
  {title} {Optical pressure sensors for high-pressure--high-temperature studies
  in a diamond anvil cell},\ }\href
  {https://www.tandfonline.com/doi/full/10.1080/08957950701659593} {\bibfield
  {journal} {\bibinfo  {journal} {High Pressure Research}\ }\textbf {\bibinfo
  {volume} {27}},\ \bibinfo {pages} {447} (\bibinfo {year} {2007})}\BibitemShut
  {NoStop}%
\bibitem [{\citenamefont {Momma}\ and\ \citenamefont
  {Izumi}(2011)}]{momma2011vesta}%
  \BibitemOpen
  \bibfield  {author} {\bibinfo {author} {\bibfnamefont {K.}~\bibnamefont
  {Momma}}\ and\ \bibinfo {author} {\bibfnamefont {F.}~\bibnamefont {Izumi}},\
  }\bibfield  {title} {\bibinfo {title} {{VESTA 3} for three-dimensional
  visualization of crystal, volumetric and morphology data},\ }\href
  {https://journals.iucr.org/paper?db5098} {\bibfield  {journal} {\bibinfo
  {journal} {Journal of Applied Crystallography}\ }\textbf {\bibinfo {volume}
  {44}},\ \bibinfo {pages} {1272} (\bibinfo {year} {2011})}\BibitemShut
  {NoStop}%
\bibitem [{\citenamefont {Hunter}(2007)}]{Hunter:2007}%
  \BibitemOpen
  \bibfield  {author} {\bibinfo {author} {\bibfnamefont {J.~D.}\ \bibnamefont
  {Hunter}},\ }\bibfield  {title} {\bibinfo {title} {Matplotlib: A {2D}
  graphics environment},\ }\href {https://doi.org/10.1109/MCSE.2007.55}
  {\bibfield  {journal} {\bibinfo  {journal} {Computing in Science \&
  Engineering}\ }\textbf {\bibinfo {volume} {9}},\ \bibinfo {pages} {90}
  (\bibinfo {year} {2007})}\BibitemShut {NoStop}%
\bibitem [{\citenamefont {Hunt}(2019)}]{hunt2019advanced}%
  \BibitemOpen
  \bibfield  {author} {\bibinfo {author} {\bibfnamefont {J.}~\bibnamefont
  {Hunt}},\ }\href {https://link.springer.com/book/10.1007/978-3-031-40336-1}
  {\emph {\bibinfo {title} {Advanced guide to {Python} 3 programming}}}\
  (\bibinfo  {publisher} {Springer},\ \bibinfo {year} {2019})\BibitemShut
  {NoStop}%
\bibitem [{doi()}]{doi.org/10.17863/CAM.112461}%
  \BibitemOpen
  \href@noop {} {\bibinfo {title} {{N}agle-{C}occo, {L}. {A}. {V}.; {S}teele,
  {J}. {M}. {A}.; {Deng}, {S}.; {Zhang}, {X}.; {Daisenberger}, {D}.;
  {Genreith-Schriever}, {A. R}.; {Saxena}, {S. S}.; {Grey, C. P.}; \& {Dutton,
  S. E.} \textit{DOI:10.17863/CAM.112461}. [data set].}}\BibitemShut {Stop}%
\end{thebibliography}%

\end{document}